\documentclass[10pt,preprint2]{aastex}

\def\rnum#1{{\uppercase\expandafter{\romannumeral#1}}}

\def\ltsim{ \,{}^<_\sim\, }
\def\gtsim{ \,{}^>_\sim\, }

\def\today{\number\year\space \ifcase\month\or
  January\or February\or March\or April\or May\or June\or
  July\or August\or September\or October\or November\or December\fi
  \space\number\day}
\def\now{\number\year\space \ifcase\month\or
  January\or February\or March\or April\or May\or June\or
  July\or August\or September\or October\or November\or December\fi
  \space\number\day .\number\time}

\slugcomment{To appear in PASP}

\shorttitle{DAOSPEC}
\shortauthors{Stetson \& Pancino}

\begin{document}


\title{DAOSPEC: an automatic code for measuring equivalent widths\\in 
high-resolution stellar spectra}


\author{Peter B. Stetson\footnote{Based on data obtained from  UVES@VLT,
GIRAFFE@VLT, HARPS@VLT located in the ESO observatory of Cerro Paranal, Cile.
Also based on data from FOCES@CAHA 2.2m telescope in Calar Alto, Spain. Data from
the VALD and GEISA online databases has been employed as well.}}
\affil{Herzberg Institute of Astrophysics, National Research Council of
    Canada, 5071  West Saanich Road, Victoria, BC V9E 2E7, Canada }
\email{Peter.Stetson@nrc-cnrc.gc.ca}

\and

\author{Elena Pancino}
\affil{INAF -- Bologna Observatory, via Ranzani 1, I-40127 Bologna,
Italy }
\email{elena.pancino@oabo.inaf.it}




\begin{abstract}  
DAOSPEC is a Fortran code for measuring equivalent widths of absorption lines in
stellar spectra with minimal human involvement. It works with standard FITS
format files and it is designed for use with high resolution (R$>$15000) and
high signal-to-noise-ratio (S/N$>$30) spectra that have been binned on a linear
wavelength scale. First, we review the analysis procedures that are usually
employed in the literature. Next, we discuss the principles underlying DAOSPEC
and point out similarities and differences with respect to conventional
measurement techniques.  Then experiments with artificial and real spectra are
discussed to illustrate the capabilities and limitations of DAOSPEC, with special
attention given to the issues of continuum placement; radial velocities; and the
effects of strong lines and line crowding. Finally, quantitative comparisons
with other codes and with results from the literature are also presented.

\end{abstract}



\keywords{Data Analysis and Techniques}


\section{Introduction}

Most of the information we acquire about the chemical compositions and
atmospheric parameters of stars comes from spectral absorption features, both
molecular bands and atomic lines. One of the most powerful and well-understood
techniques to disentangle the chemical abundance from, for instance, the effects
of temperature and gravity on absorption features employs measurements of
equivalent widths (EWs).

Usually, chemical abundances are best derived when the EW analysis is based upon
high-resolution (R$=\lambda / \delta \lambda \gtsim$15000) spectra with
relatively high values of the signal-to-noise ratio (S/N$\gtsim$30). This level
of data quality was typical of the best that could be achieved in the 1970s and
now it has become a sort of lower limit for the standard\footnote{Typically,
R$\simeq$40000 and S/N$\simeq$50--100 are used in the optical.} quality expected
for deriving accurate chemical abundances from optical spectra ($\lambda \sim
3200$ -- 10000$\,$\AA). Recently, it has become possible to achieve similar data
quality (R$\simeq$10000, S/N$\simeq$30) in the near infrared wavelength range
(up to $K$-band wavelengths, $\lambda \simeq 2.2 \mu$m). 

As technology has advanced rapidly in the last few decades, telescopes of
increasingly larger aperture have become available, as well as spectrographs that allow us
not only to reach very good resolution and flux fidelity, but also to observe many objects
simultaneously. It has now become commonplace to work on sets of hundreds of
high-resolution spectra from fiber spectrographs such as, e.g., FLAMES mounted
at the European Southern Observatory Very Large Telescope in Chile
\citep{flames}. Our knowledge of atomic constants for many of the absorption lines
present in stellar atmospheres has also advanced and we are now typically able to use
several tens of lines for a few dozens of different chemical species.

The actual process of measuring EWs has until recent years mostly been tackled with the
help of interactive graphical software, such as IRAF\footnote{Image Reduction
and Analysis Facility. IRAF is distributed by the National Optical Astronomy
Observatories, which is operated by the association of Universities for
Reasearch in Astronomy, Inc., under contract with the National Science
Foundation} or MIDAS \citep{midas}. However, this procedure has two main
drawbacks: {\em (i)} lines are usually identified by the user in a
non-automatic way, and the measurement of each individual line requires---to say the
least---several seconds of direct human involvement (lack of {\em performance});
{\em (ii)} a certain degree of subjectivity is involved in the measurement
process, particularly due to the continuum placement by the user around each line;
this means that different users could produce different results from the same
data (lack of {\em
robustness}).  It may also be noted that the tedium of the process itself often
discourages a thorough testing of the procedures via repeated measuring
under different sets of assumptions and protocols.

Therefore, given the vastly increased volume of data and the better data quality
available these days, the community has started writing \hbox{(semi-)}automatic codes
that improve both the performance of the measuring process and the robustness of
the results. Most of these codes are not really public, in the sense that they
are designed to meet the specific goals of some research group and are
provided on a personal basis to close collaborators. 

DAOSPEC is a Fortan code to measure EWs of atmospheric absorption lines in
high resolution spectra. Our goal is to provide the performance required by the
increasing amount of data available; the convenience and repeatability of a fully automatic
routine; and availability to the whole astronomical community. With the help
of simple shell scripts, detailed in the DAOSPEC Cookbook \citep{cookbook}, a
large number of stars with different characteristics and configurations can
be processed in a fully automatic and unsupervised way. Effort has been put
into making the code compatible with standard astronomical image formats such
as the FITS standard \citep{fits0,fits1,fits2,fits3} and to use standard Fortran
coding and libraries. 

One advantage of a fully automatic code that is also publicly available is that
the results are reproducible and testable in variety of conditions of use and by
many different users. Indeed, such a set of independent tests is presently under
way in the small but growing community of DAOSPEC users (see also
Section~\ref{sec-comp}). 

In Section~\ref{sec-ew} below we describe the EW measurement process in general,
and its implementation in a few public software packages; in Section~\ref{sec-dao}
we give a general description of DAOSPEC itself; in the various subsections, we
discuss in depth a few topics which are relevant to the most common scientific
cases, including  radial-velocity determination (Section~\ref{sec-rad}); continuum
placement (Section~\ref{sec-con}); strong lines and blended lines
(Section~\ref{dao-ew}); uncertainty estimates (Section~\ref{dao-err}); and
dependence of the results on the choice of input parameters
(Section~\ref{dao-pars}) and on spectral quality (Section~\ref{dao-qual}). Some
performance considerations are also presented (Section~\ref{dao-perf}). Comparisons
with other codes and with previously published measurements are presented and
discussed in Section~\ref{sec-comp}. Finally, in Section~\ref{sec-dis} we summarize
our conclusions.

\section{EW Measurements in General}
\label{sec-ew}

\begin{figure}[t]
\includegraphics[angle=270,scale=.28]{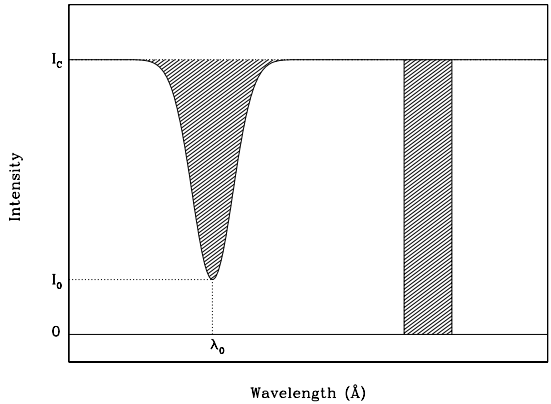}
\caption{Classical definition of EW: the width of a rectangular line
(right) with the same area as the actual absorption line (left). 
See text for details. I$_C$ is the continuum level while I$_0$
is the specific flux at the line center $\lambda_0$, measured upward from
zero.\label{ewdef}}
\end{figure}

Figure~\ref{ewdef} illustrates how the EW of a spectral line is defined,
starting from the continuum level (I$_C$) and the specific flux at the line
center (I$_0$). The EW represents the strength of an absorption line, and it is
defined as the width, in wavelength units, that a rectangular stripe  of height
I$_C$, in intensity units, would have if it had the same area as the actual line
or---in other words---the wavelength interval that would be covered by a
hypothetical, perfectly opaque absorption feature that removes the same amount
of energy from the continuum flux. More practically, for an isolated spectral
line in a discretely-sampled spectrum from a real instrument:

\begin{equation}
\label{eq-sum}
EW = \Delta \lambda \sum _i \frac{I_{C_i}-I_i}{I_{C_i}}
\end{equation}

\noindent  
where $\Delta \lambda$ is the (constant) pixel size, I$_{C_i}$ is the continuum
level at the wavelength of the $i$-th pixel, and I$_i$ is the actual flux
received by the $i$-th pixel.

In a real spectrum, however, noise, spectral defects and neighboring lines can
affect the perceived profile of an absorption line, and an EW is not necessarily
readily measured as a direct summation of the pixel fluxes. Since the line
profile is generally dominated by the point-spread function of the slit image,
which can often be approximated by a Gaussian profile\footnote{Different line
profiles (i.e., Voigt, Lorentz) can be adopted for lines that significantly
deviate from the Gaussian form.}  (see Section~\ref{dao-ew}), often a {\it
numerical fit\/} of a Gaussian-shaped function $g(\lambda)$ is adopted instead
of the direct numerical summation:

\begin{equation}
\label{Gauss}
g(\lambda) = A~e^{-(\lambda - \lambda_{0})^2/2\sigma^2}, A > 0
\end{equation}

\begin{equation}
\label{Gauss2}
I_i = I_{C_i} \left[1 - g(\lambda_i)\right]
\end{equation}

\noindent 
(i.e., the specific flux $I_i$ assumes a minimum value $I_0 = I_{C_I}-A$ at
$\lambda_i = \lambda_0$, and $I_i \rightarrow I_{C_i}$ for $\lambda_i$ far from
$\lambda_0$).  The estimated EW thus becomes

\begin{equation}
\label{eq-fit}
EW = \sum _i \frac{I_{C_i}-I_i}{I_{C_i}} \Delta \lambda \rightarrow \int 
g(\lambda) \lambda
\end{equation}

\noindent 
where $\Delta \lambda$ is the step of the integration process that can be made
very small ($\rightarrow d \lambda$).

No measurement means much without an associated uncertainty. When fitting
lines to obtain EWs with Equation~\ref{eq-fit} in a digital computer, a standard error 
can be computed for each line with various mathematical techniques that
estimate the goodness of the fit (e.g., least-squares formal errors). These
confidence intervals can also take into account the quality of the spectrum (R, S/N, 
and---to a certain extent---crowding and spectral defects, etc.).  
\citet{cayrel} gives
approximate formulae for estimating the uncertainty of an EW as a function of spectral
quality:

\begin{equation}
\label{eq-cayrel}
\delta EW \simeq 1.5 \frac{\sqrt{\Delta \lambda \cdot FWHM}}{S/N} 
          \simeq 1.6 \frac{\sqrt{\Delta \lambda \cdot EW}}{S/N} 
\end{equation}

\noindent 
where FWHM is the the full width at half maximum of the line. (For a perfect
Gaussian profile, FWHM=$2.355~\sigma$.) Of course, as noted in the paper by
\citet{cayrel}, this is a lower limit to the actual uncertainty because the true
continuum level is never perfectly known, but it too has an associated
uncertainty that should be propagated through the whole calculation.  However,
the effect of continuum placement (see below) is very difficult to take
completely into account.  

Ways to estimate the uncertainty due to the continuum placement in each pixel have been
obtained by \citet{ewdet}, \citet{bohlin} and \citet{lev02} among others. The
idea is that the derivative of Equation~\ref{eq-fit} has to be computed not only
with respect to $I_i$ but also with respect to the continuum level in each
pixel, $I_{C_i}$, and the total uncertainty then becomes\footnote{For those
programs that evaluate the continuum {\it locally} in two windows beside each
line, the full uncertainty can also be estimated with the alternate method
provided by \citet{bohlin}.}

\begin{equation}
\label{eq-err}
\delta EW = \sqrt{\left(\sum _i (\delta I_i)^2 \frac{\partial EW}{\partial I_i}\right)^2+
                  \left(\sum _i (\delta I_{C_I})^2 \frac{\partial EW}{\partial I_{C_i}}\right)^2}.
\end{equation}

\noindent     
where $\delta I_i$ and $\delta I_{C_i}$ are the uncertainties on  $I_i$ and
$I_{C_i}$, respectively; e.g. these would be the Poisson errors $\sqrt{I_i}$
and $\sqrt{I_{C_i}}$ if the spectra are in units of raw detected
photoelectrons. Otherwise, of course, the errors must  be appropriately adjusted
for any renormalizing of the data. As pointed out by \citet{lev02}, the above
formula still provides only an estimate of $\delta EW$ because when the spectra
are resampled to constant wavelength bins, the errors in the single pixels
become correlated. The correlation term that should be added to
Equation~\ref{eq-err} appears to be small, and as we will see in
Section~\ref{sec-con}, there are other uncertainty sources that dominate the
continuum error budget, such as line crowding and blanketing by very weak
lines.  

\begin{figure}[t]
\includegraphics[scale=0.38]{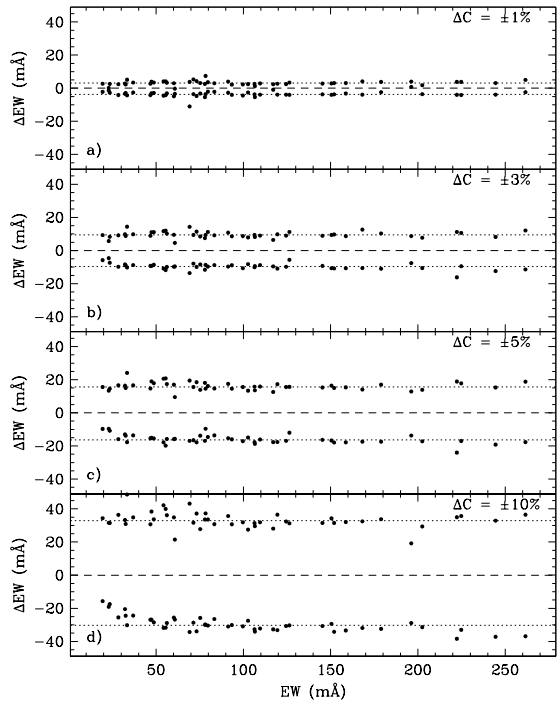}
\caption{Differences between EW measured on the adopted ``best" continuum level
and the EW of the same lines measured on a continuum level altered by $\pm$1\%
(Panel a), $\pm$3\% (Panel b), $\pm$5\% (panel c) and $\pm$10\% (Panel d). The
spectrum is that of star 2129 in the open cluster Cr~110 (Pancino et al., in
preparation) \label{fig-levs}}
\end{figure}

A very simple experiment shows the impact of continuum placement on the EW
measurements. An observed  spectrum of star 2129 in the open cluster Collinder~110
(Pancino et al., in preparation) was normalized to an arbitrary ``best" continuum
in the range 5500--6500~\AA\  and the EW of $\sim$50 lines were measured with the
task {\em splot} in IRAF. This spectrum has a resolution of R$\simeq$30000 and a
S/N ratio of approximately 75 (per pixel). Later, the lines' EWs were remeasured
adopting a continuum level higher or lower than the adopted ``best" continuum by
1\%, 3\%, 5\% and 10\%, respectively. The resulting $\Delta$EW values are shown in
the four panels of Figure~\ref{fig-levs}. As can be seen, except for a few lines with the smallest
EWs, all measured line strengths are altered by approximately constant values:
$\sim\pm$3--3.5~m\AA\  in the case of a $\pm$1\% misplacement,
$\sim\pm$9.5~m\AA\  in the $\pm$3\% case, $\sim\pm$16~m\AA\  in the $\pm$5\%
case, and $\sim\pm$30--32~m\AA\  in the pessimistic case of a 10\% misplacement.
Of course, noise, resolution, line crowding and line blanketing also have an
impact on the actual $\Delta$EW values, but the above experiment gives a
quantitative idea of what happens when the continuum level is badly misestimated (see
also Section~\ref{sec-conterr}).

In the following subsections we examine the various operational steps that must
be undertaken when measuring the EW of a line. We will provide practical
examples based on a few publicly available programs. The IRAF task {\em
noao.oned.splot} is taken as a good representation of those graphical
interactive packages that allow the user to manually\footnote{When dealing with
many spectra having similar characteristics, many users develop their own
pipelines in the IRAF script language or the MIDAS equivalent, but these
pipelines cannot really be considered here, since they are often highly specific
to the problem treated and are seldom made public.} measure EWs from their
spectra; we believe that inferences drawn from the IRAF package will apply to
other analogous routines, such as that provided within MIDAS, and others. The
Fortran program SPECTRE\footnote{http://verdi.as.utexas.edu/spectre.html}
\citep{spectre} is taken as an example of a more automated routine and since it
is part of the widely used MOOG
package\footnote{http://verdi.as.utexas.edu/moog.html} \citep{moog} it serves as
a very good reference in the following discussion. ARES \citep{ares} and EWDET
\citep{ewdet} are  two programs written in C++ and Fortran respectively.
Although not entirely objective (due to user involvement in the continuum
placement, see Section~\ref{placement}), to our knowledge they are the only
fully automated codes publicly available\footnote{We are aware of the existence
of other codes that are not really public, and therefore are not considered in
the following discussion. For instance, one such program, ROSA, has been written
by R.~Gratton (E.~Carretta, private communication) and another, {\em fitline},
by P.~Fran\c cois (P.~Fran\c cois, private communication).}.

\subsection{Line Finding and Identification}
\label{int-find}

Within interactive tasks such as {\em splot} or SPECTRE, the processes of line
{\em finding} and {\em identification} are generally joint and happen in the
same initial phase of the measurement. By {\it line finding\/} we mean the
process of specifying where in the observed spectrum are perceived dips representing good candidate
absorption lines. By {\it line identification\/}, on the other hand, we mean the
process of specifying which of the perceived absorption features corresponds to a
particular atomic transition. The latter task requires some knowledge of the
radial velocity of the star to infer the rest wavelength of the line for
comparison with laboratory line lists.

Within IRAF, {\em splot} requires the user to mark with the cursor, on a 
radial-velocity-corrected spectrum, the continuum position on both sides of a line.
Apart from helping to define the continuum level, this also tells the program the
line's approximate location ({\em initial finding}). The program then fits
the appropriate function and determines a precise wavelength for the line center ({\em
refined finding}), as well as the EW. Finally the user confirms whether the
measured line center is acceptably close to the laboratory wavelength ({\em identification}). 

SPECTRE also needs a radial-velocity-corrected spectrum as input, and accepts
entry of the laboratory wavelength of a line (or reads a list of wavelengths
from a file).  The measuring process {\it per se\/} has a similar structure to
that of IRAF: {\em initial finding} based upon the rest wavelength, {\em refined finding}
(and fitting) of the perceived absorption feature closest to the rest wavelength and final {\em
identification} by decision of the user. In the easiest cases, users just need
to provide SPECTRE with the spectrum and a line list, and to check subsequently whether the
program has done a good job.  Thus, the procedure can be much faster than with
IRAF. However, SPECTRE also provides several interactive tools to improve the fit in more
difficult cases, such as continuum adjustments, the use of partial line
profiles, and others.

In ARES and EWDET, {\em line finding} is completely automatic and independent from
the {\em identification} procedure.  

ARES uses local maxima of the second derivative---defined as zeros of the third
derivative---to identify the centers of absorption lines. In the paper that
presents the code, \citet{ares} have shown (see, e.g., their Fig.~3; see also
Section~\ref{sec-fit} below) that this approach can enhance the detection of
highly blended lines not readily apparent in the original spectrum.  However,
the effect of noise dramatically propagates in the numerical derivatives even
for spectra with high S/N ratios. Therefore, in practice the observed spectrum
is artificially smoothed to reduce confusion between noise and real features
before computing the numerical derivatives. This is especially important since
the FWHM is left as a free parameter (see Section~\ref{sec-fit}) and cannot
therefore be used to distinguish between an absorption line and a noise
artifact. To avoid further problems, the input parameter {\em lineresol} tells
the code the minimum distance between lines in \AA ngstr\"oms.  ARES requires
that the spectrum be pre-corrected for the radial velocity to identify detected
lines with entries in the input line list on the basis of their rest
wavelengths. 

In EWDET, the spectrum is examined, pixel by
pixel, to find all the points that deviate from a previously found continuum by
more than 2$\sigma$. Those points are further examined to see whether they have
neighbors that also deviate from the continuum, and if so, the line {\em finding}
is considered complete. The {\em line identification} is left to the user in the case of
EWDET, and, to facilitate this task, the radial velocity is used as an input to
compute the rest wavelength of each line. 

One side effect of all these methods, which can considerably slow down the
procedure, is that the radial velocity of each star must be known {\em a priori}.
In some cases, as we have seen, the input spectrum must already be shifted to zero
radial velocity.  

\subsection{Continuum Placement}
\label{placement}

Continuum placement is a source of uncertainty for any EW measurement, and there
is as yet no generally accepted best practice in estimating the continuum level
in a stellar spectrum densely populated with absorption lines.  For instance,
when writing a program for extracting EWs, one must consider whether to use a
{\em local} or {\em global} continuum normalization, or a combination of the
two.  The approach employed in DAOSPEC will be examined in depth in
Sections~\ref{sec-dao} and \ref{sec-con}.  Here, we describe the procedures
adopted by SPECTRE, {\em splot}, ARES and EWDET.

SPECTRE requires spectra that are already roughly normalized to their continua.
This is necessary because the user can only, if needed, re-adjust the local
continuum {\it level\/} and not its global {\it shape\/}. Therefore it is
optimal to normalize spectra that have steep or strongly curved continua before
feeding them to SPECTRE. 

IRAF/{\em splot} allows the user to mark separate left and right continuum
levels relative to the chosen line, and these can have very different values. 
However, since the continuum level {\it within\/} the absorption line is
evaluated by interpolating a straight line between the two continuum samples, a
preliminary continuum correction is again desirable in the case of strongly
curved or steep spectra. At least one of the private packages of which we are
aware (E.~Carretta, private communication) also chooses to normalize the
continuum based on windows on the short- and long-wavelength sides of each
absorption line. So, this package, like both SPECTRE\footnote{We point out that
SPECRE provides additional tools that help continuum fitting, such as the one
colloquially known as {\em boinking}.} and {\em splot}, therefore requires (or
gives better results after) a {\em global} continuum normalization, i.e., the
procedure of dividing the whole spectrum by a fitting function (usually a
polynomial or a spline function), followed by minor {\it localized\/} continuum
adjustments.  

ARES also finds the continuum level locally around each line, using a $\kappa\sigma$
rejection algorithm. The parameters of the fit and of the $\kappa\sigma$ rejection,
however, are highly customizable and flexible. This is the only point in ARES
where the user has a strong impact on the outcome: trying many different
combinations of parameters is important for guaranteeing  a good result, and an
inexperienced user can obtain worse results than an experienced or extremely
careful user, as can also happen with {\em noao.oned.continuum}. Moreover, the
best combination of parameters will change for different sets of spectra, and
various users will have different opinions on how the continuum should be placed.
In summary, continuum placement is the main reason why the results obtained with
ARES cannot be considered entirely objective. 

EWDET chooses a {\em global} continuum normalization approach: the continuum
determination is done by automatically fitting a curve to the whole spectrum
with several iterations of point rejection above and below sigma levels
specified by the user. This is not significantly different from what is
done in the IRAF task {\em noao.oned.continuum}, and as in the case of ARES, it
introduces a certain degree of subjectivity in the process. EWDET however, unlike
other programs, stores in its memory the uncertainty of the continuum placement and
uses it in the computation of the final EW uncertainty (see Section~\ref{sec-err}).

The choice of the continuum normalization method is crucial and will be
discussed further in Section~\ref{sec-con}.

\subsection{Line Fitting}
\label{sec-fit}

IRAF offers the widest range of possibilities: a line can be measured by summation of
the pixel values (i.e., without any fitting procedure, Equation~\ref{eq-sum}), or by
fitting a Gaussian function (Equation~\ref{eq-fit}), a Voigt profile, or a Lorentz
profile. The task can measure both absorption and emission lines. Finally a deblending
routine is also available, where the user interactively marks the region to
investigate, the peaks of the blended lines, and so on. SPECTRE offers similar
options, although it is more specifically designed for the classical analysis of
isolated absorption lines. ARES and EWDET use Gaussian profiles, performing fits
similar to our Equation~\ref{eq-fit} and reporting the fit parameters in their output
files. ARES, however, appears to be able to fit a line with multiple Gaussian
functions to improve the fit of each line profile, and the number of Gaussians
actually used is listed in the output file. 

Thus, the most widely used profile is a Gaussian function. Generally speaking,
three parameters are free in the fit: the centroid $\lambda_0$, the standard
deviation  $\sigma$ or, equivalently, the full-width at half-maximum, and the
depth of the line center ($A$ in the notation of our Equation~\ref{eq-fit}). Of
these three, here we consider the FWHM. In
an observed spectrum, the line profile is a convolution of the instrumental
profile with an intrinsic line profile. At the resolution that is typically used
for routine abundance analysis (i.e., from R$\sim 20,000$ to R$\sim50,000$), the
line profile is still dominated, for most lines, by the instrumental profile,
which either is the same for all lines, or scales as the wavelength when
multiple echelle orders are patched together. Therefore, in principle, there is
no need to leave this parameter free for each individual line once the
instrumental characteristics have been determined.  

Most existing packages, however, choose to leave this parameter free for each
line, as is done in {\em splot}, SPECTRE, ARES, and EWDET. As an advantage, the
derived FWHM can be used {\em a posteriori} to indicate whether a line has been
properly measured: lines with a FWHM significantly larger than average are most
probably unresolved blends, while lines with a FWHM significantly smaller than
average could be noise spikes, telluric features, or spectral defects. Another
positive side of this approach is that it compensates, to some degree, for
saturation. When a line becomes strong enough to just enter the saturation regime,
it can still be mimicked with a slightly wider Gaussian, with a modest error in
the EW measurement (of course, very strong lines cannot be correctly reproduced
this way.) The main disadvantage is allowing the FWHM to be completely free
complicates the deblending capabilities of an automated program when the noise
level in the spectrum is appreciable.  Most obviously, an apparently broad
feature could be modeled by a single profile with a large FWHM or by multiple
narrower profiles.  Moreover, even when the number of components in a blend is 
considered known, numerical degeneracies between the widths and strengths of the
various overlapping features increase the uncertainties in the fitting parameters that
are produced by the analysis.

\subsection{Uncertainty Evaluation}
\label{sec-err}

Uncertainties in the measurement of EWs are often neglected when deriving
chemical abundances. Often, having measurements of $n$ lines for some chemical
element, it is convenient to compute the mean and variance of the single-line
abundances and to use $\sigma / \sqrt{n-1}$ as the uncertainty of the net
abundance adopted for that particular element.  Using a good estimate of the EW
uncertainty of each line to perform a weighted average would give somewhat
better results, but the dominant source of uncertainty in the adopted chemical
abundances is often the uncertainty in the stellar atmospheric parameters.  This
explains at least in part why little effort is sometimes put into evaluating EW
uncertainties (see also Section~\ref{sec-ew}).  

Still, when writing a package that measures EWs, individual uncertainties can
easily be estimated, and indeed they are of fundamental importance in particular
cases, e.g., temporal variations of the EW. In the case of {\em splot}, each of
the pixels used in the fit is assigned its own standard error through the
construction of a statistical noise model based on Poisson photon noise (readout
noise is usually negligible for spectra with high S/N ratios).  The error
estimate, however, is optional and can only work if the user inputs an
appropriate gain value  if the spectral intensities are recorded in some
instrumental  data-number units rather than in raw detected photoelectrons.
Furthermore, if the input spectrum has been prenormalized with a flat-field
spectrum or to a provisional continuum, the numerical  conversion from Poisson
errors to intensity-unit errors will be a  function of wavelength. In SPECTRE
and ARES, no error computation routine is provided, for the reasons outlined
above. EWDET, on the other hand, performs a full error computation that is
described in detail by \citet{ewdet}. In short, each pixel used in the fit is
assigned its own uncertainty estimate based on propagation of errors through the
Gaussian fitting procedure\footnote{The Gaussian fit used by \citet{ewdet} is
different from our Equation~\ref{eq-fit}, in that the formula by \citet{ewdet}
is strictly valid only when I$_C$=1, while our formula is valid in the general
case, even if I$_C\neq$1.}, including the continuum placement uncertainty, which
is almost never considered in other packages. Pixel uncertainties are then
summed in quadrature after multiplication by the pixel size.  

\begin{figure*}[t]
\includegraphics[scale=0.93]{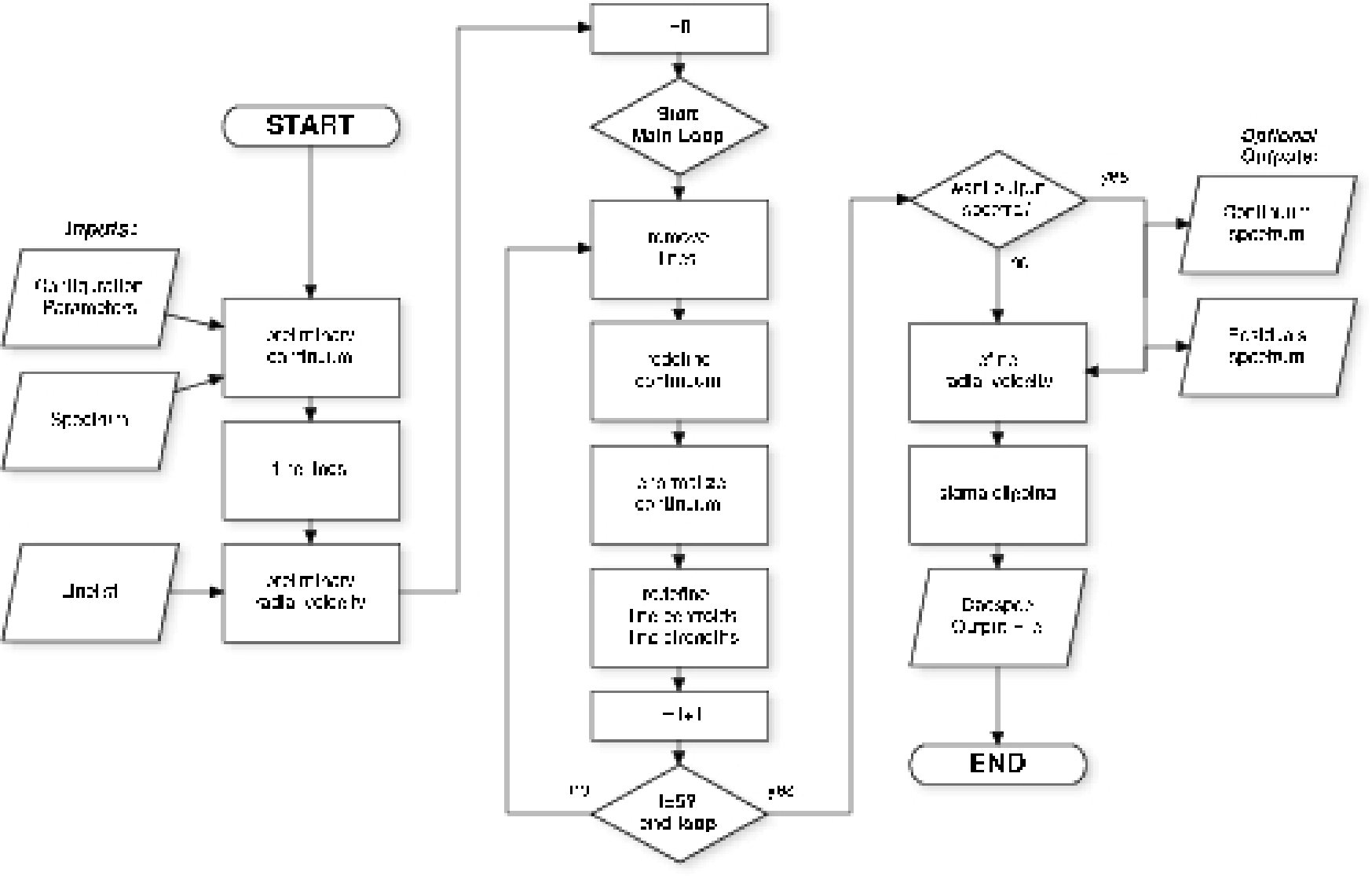}
\caption{Flow chart diagram for DAOSPEC. \label{flow}}
\end{figure*}

\section{DAOSPEC}
\label{sec-dao}

DAOSPEC\footnote{\tt http://cadcwww.dao.nrc.ca/stetson/daospec/ \\
http://www.bo.astro.it/$\sim$pancino/projects/daospec.html} is a code written in
standard Fortran 77. The version that we have made generally available  makes
use of the cfitsio libraries\footnote{\tt
http://heasarc.nasa.gov/docs/software/fitsio/fitsio.html} \citep{cfitsio} for
dealing with standard FITS-format spectra.  IRAF format (.imh, .pix) files can
also be used, and we have included an optional graphical display\footnote{``SuperMongo:'' \tt
http://www.astro.princeton.edu/$\sim$rhl/sm/} that allows the user to monitor
the progress of the analysis.  However, none of these should be considered
an integral or essential part of the algorithms we discuss here.  Practical and detailed
help on how to obtain, install, and use DAOSPEC can be found in the DAOSPEC
Cookbook\footnote{\tt http://cadcwww.dao.nrc.ca/stetson/daospec/daospec.ps \\
http://www.bo.astro.it/$\sim$pancino/docs/daospec.pdf} \citep{cookbook}, or
by directly contacting the authors.

We briefly describe here the general code structure depicted as a flow
chart in Figure~\ref{flow}. More details on the most important steps are given
in the following subsections.  The reduction path consists of three main stages:
{\em (i)} the input and preparation phase, shown on the left side of
Figure~\ref{flow}; {\em (ii)} the main iteration loop, center of
Figure~\ref{flow} and {\em (iii)} the evaluation and output phase, right side of
Figure~\ref{flow}.  

Three inputs, shown on the left side of Figure~\ref{flow}, are needed: {\em (i)}
a small set of configuration parameters, specifying among other things the wavelength
limits of the spectral region of interest, a first guess at the FWHM, the order of the polynomial
to be used for the continuum normalization, and other basic information required
by the code; {\em (ii)} the reference line list, which contains the laboratory
wavelengths of as many clean and unblended lines as possible, and may contain
any additional textual information---such as atomic parameters and notes---that
the user finds convenient; and {\em (iii)} a spectrum in standard FITS or IRAF
format, which must be binned linearly with wavelength.  Items {\em (i)} and {\em
(ii)} are specified once at the very beginning of the reduction process; any
number of individual spectra {\it (iii)} may be then analysed one after another,
at the user's pleasure.  A more detailed description of this procedure is given
in the Cookbook.  

With the above inputs, the code performs an initial continuum fit with the help
of Legendre polynomials of the order specified by the user, and performs the line
finding procedure (see also Section~\ref{sec-rad}). Then, a preliminary
correlation of detected lines with the input list of laboratory wavelengths is
performed, to derive an initial radial-velocity estimate. 

Now the main loop starts (center of Figure~\ref{flow}). The detected lines are
provisionally subtracted from the spectrum and the remaining residual spectrum
is used to refine the continuum normalization (see also Section~\ref{sec-con})
by means of robust non-linear least squares. With the newly normalized spectrum,
the individual line centroids and strengths are then refined, along with a value for the FWHM
which is either constant for all lines, or considered to be a linear function of
wavelength (see also Section~\ref{dao-ew}).  This whole procedure is repeated
five times.  (The number ``five" was chosen arbitrarily and---so far---it appears to
be adequate.  It can be changed trivially if future experience indicates a need.)

After final execution of the refinement loop, the fitted wavelengths
of the individual lines are compared to the table of laboratory wavelengths.  A
simple outlier-clipping algorithm is used to determine the final radial-velocity
estimate for the star and, at the same time, to specify which of the detected lines lie
sufficiently close to tabulated laboratory wavelengths to be regarded as 
legitimate detections of particular atomic transitions.  Equivalent widths
are computed from the fitted line parameters, and uncertainties in those EWs
are determined from the residual noise remaining in the spectrum within the
wavelength range of the line profile (see also Section~\ref{dao-ew}).  

The code output is a text file containing in its header the estimated FWHM at
the center of the spectrum; the estimated radial velocity of the star with its
standard error; the number of lines used to detemine the velocity; and the
root-mean-square value of the pixel-by-pixel flux residuals remaining in the
spectrum after subtraction of all the fitted lines, expressed as a percentage of
the fitted continuum flux.  Following that, the file contains a list of all the
lines found by DAOSPEC having strengths greater than a user-specified minimum
EW.  Both their observed and inferred rest wavelengths are reported, these
latter computed on the basis of the estimated radial velocity. If a particular
detected line has been identified with an entry in the input line list, the
laboratory wavelength of the line along with any additional textual information
from the input line list is also reproduced in the output. 

Optionally, (right side of Figure~\ref{flow}), the code can be instructed to
produce two FITS- or IRAF-format spectra, one containing the final estimated
polynomial continuum model used to normalize the spectrum, and another
containing the spectrum of the flux residuals, of the flux residuals, after the
spectrum has been divided by the fitted continuum and all the lines found by 
DAOSPEC have been fitted and digitally subtracted.

\subsection{Line finding, line identification and radial velocity}
\label{sec-rad}

Line finding in DAOSPEC is performed with the help of a tuned second-derivative
filter to identify local minima. The user-specified initial guess of the FWHM is
used to optimize the distinction between valid lines and noise or continuum
features\footnote{This means that DAOSPEC is effective in separating noise
from real spectral features as long as the FWHM of a line is adequately sampled,
i.e., by at least 2 pixels.}.  
During the main iteration loop (see Figure~\ref{flow}) the initial FWHM estimate
is refined at the same time as the line centroids and strengths, by the method
of robust non-linear least squares, and provisional line detections are
abandoned if refined EW estimates indicate they are too weak to be of interest. 
Note that, as mentioned above, the user may specify whether the FWHM is to be
regarded as constant over the entire wavelength range of the spectrum, or as
proportional to wavelength.

Preliminary line identification is performed by a simple cross-correlation of the
detected lines with the input list of laboratory wavelengths:  that wavelength
shift that corresponds to the greatest number of matching lines is adopted as
indicating the preliminary radial-velocity estimate for the star.  (The user can
help make this process reliable by supplying as {\em complete} and {\em
appropriate} a line list as possible for the anticipated spectral class of the star,
and for the presumably known wavelength range, resolution, and S/N ratio of the
spectra.) After completion of the main refinement loop, these cross-identifications
are reconsidered on a line-by-line basis. A particular cross-identification of
a detected line with a laboratory transition is regarded as valid if the
radial velocity implied by that provisional identification agrees with the adopted
final radial velocity of the star within a user-specified tolerance.  Therefore,
once the laboratory line list and the significance and identification tolerances
have been specified by the user, both the line {\it finding\/} and {\it
identification\/} procedures are completely automated, and the observed radial
velocity of the star is one of the program {\em outputs}, not {\em inputs}. This
has an impact on performace, as discussed in Section~\ref{int-find}.

The final radial-velocity measurement is provided together with its line-by-line
variance, ($\sigma$), and the number of lines ($n$) actually used in its
determination. These can in principle be combined to provide a final error
estimate as $\sigma / \sqrt{n-1}$. However, when measuring spectra that have
not been specifically intended for high-precision radial-velocity measurements,
special care must be taken. For example, if there are thermal changes between
daytime lamp observations and night-time star observations, or if the star image
can be off-center in a finitely wide spectrograph slit, there may well be
systematic velocity errors that are not reflected in the line-by-line velocity
variance.

\begin{figure}[t!] 
\includegraphics[scale=0.35]{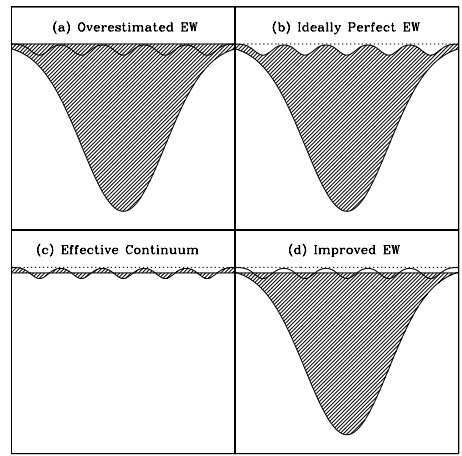}
\caption{Effect of different choices of the continuum (see text for details).
Panel (a): A strong spectral line is sketched as a Gaussian; small lines and
spectral defects (small Gaussians) pollute the line; if the EW is measured with
respect to the {\em true continuum} (horizontal line), it will be overestimated
because the flux absorbed in the weak lines will be added to the flux absorbed
in the strong line. Panel (b): as before, but now the shaded area represents the
true amount of flux absorbed in the strong line alone, exclusive of the flux
removed from the spectrum by the weak lines. Panel (c): here the strong spectral
line has been fitted and removed, and the {\em effective continuum} (solid line)
is obtained by balancing to zero the residuals (dashed area), which include
noise (not represented here for clarity) and all the small polluting features
(small Gaussians); the {\em effective continuum} (dotted line) is therefore
lower than the {\em true continuum}. Panel (d): the EW is now estimated with
respect to the {\em effective continuum}, a depressed continuum that includes,
statistically, the effect of typical local pollution by weak lines; the EW
estimate is thus improved. \label{continuum}}
\end{figure}

There is the possibility of correcting for these effects if at least one of the
telluric absorption bands\footnote{We remind the reader that while night-sky
{\em emission} lines can be used to identify and correct thermal or other changes in
spectrograph alignment, they {\it cannot\/} be used for correcting  errors due to
mis-centering of the star in the slit since the sky illuminates the slit
uniformly.} of O$_2$ or H$_2$O is included in the observed spectral range. A very
accurate list of rest wavelengths of atmospheric absorption lines can be obtained
from the GEISA\footnote{Gestion et Etude des Informations Spectroscopiques
Atmosph\'eriques, {\tt http://ara.lmd.polytechnique.fr/
htdocs-public/products/GEISA/HTML-GEISA/}} database \citep{geisa1,geisa2}. A
separate measurement can be performed using such a telluric line list as input and
forcing DAOSPEC to explore a limited radial velocity range close to 0~km~s$^{-1}$.
Our experience on red giants in $\omega$~Centauri\footnote{Using ESO GIRAFFE data
with R$\simeq$20000 and S/N$\simeq$50--100, and approximately 150 clean and
unblended lines around the 6300\AA\  region.} \citep{p07} and in open
clusters\footnote{Using Calar Alto FOCES data with R$\simeq$30000 and
S/N$\simeq$50-100, and a few hundred clean and unblended lines.} (Pancino et al.,
in preparation) shows that  the radial velocity shifts measured in the telluric
absorption lines can be up to 1--3~km~s$^{-1}$ and, what is more important, the
associated uncertainty can be of order 0.5~km~s$^{-1}$.  However, when these
measurements are used as {\it corrections\/} to the stellar radial velocities,
formal errors of 0.1--0.3~km~s$^{-1}$ can be easily achieved.

\subsection{Continuum Normalization}
\label{sec-con}

What is generally understood as the {\em continuum} of a spectrum is the level of flux
that one would have if {\it only\/} continuum-opacity sources---i.e, no discrete absorption 
lines or bands---were present: we will call
this the {\em true continuum}. Traditionally we seek the true continuum to anchor our
fitting function when measuring the EW of a particular line
(Equations~\ref{eq-sum} and \ref{eq-fit}). This is appropriate when the spectra are not
crowded and the line in question is not contaminated by neighboring lines.
Otherwise, true continuum determinations are particularly difficult and in most
cases it is also very difficult to state the associated uncertainty. Moreover, we contend that
when measuring lines in crowded spectra, what we really want to determine is
{\em not} the true continuum level at all.

In fact, we are trying to estimate how much flux one particular atomic transition
is subtracting from the net output flux of the stellar atmosphere. But we need to separate that from
the flux subtracted (or added) by other contaminating effects, such as: {\em
(i)} other atomic or molecular lines that are too weak or too close to our line
to be distinguished; {\em (ii)} the noise, which for an infinite number of pixels
should give an average zero contribution, but for a small number of pixels could
give a non-zero contribution; {\em (iii)} detector defects; and {\em (iv)} artifacts
from the imperfect removal of cosmic rays, sky emission lines and telluric absorption
features. Therefore, even if we were able to find the true continuum, when
evaluating the area enclosed below the true continuum, we would not be measuring
the flux of the intended transition alone, but its sum with all these other
effects (Figure~\ref{continuum}, Panels (a) and (b)). 

\begin{figure}[t!]
\includegraphics[angle=-90,scale=0.30]{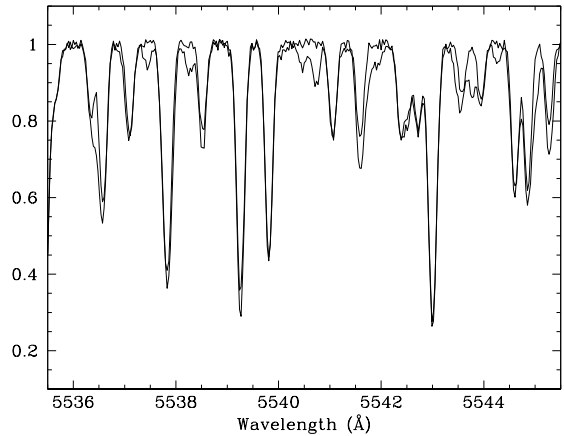}
\caption{Artificial spectra used for continuum placement
experiments, the {\em clean spectrum} (thin line) that includes only
lines with EW between 20 and 100~m\AA, and the {\em full spectrum}
(thick line), that contains also several hundreds of small lines
with EW$<$20~m\AA, the vast majority having EW$<$1~m\AA.
\label{arti}}
\end{figure}

Rather than trying to find where the true continuum lies, DAOSPEC looks for the
{\em effective continuum}, a depressed continuum that takes into account, in a statistical sense, the
unrecognized flux deficits and excesses due to all the sources described
above\footnote{This is probably the reason why the {\em effective continuum}
determined by DAOSPEC is sometimes lower than then continuum estimated by other programs,
especially in the case of crowded spectra. Note however, that programs that rely
on an automatic and {\em global} continuum placement procedure, like EWDET, tend
to be always in good agreement with DAOSPEC, placing the continuum lower than
{\em local} continuum estimates.}. Under the assumption that the unidentified and unrelated
opacity sources underneath the spectral line in question are statistically
similar to those lying at nearby wavelengths, DAOSPEC finds the effective local continuum by
robustly balancing to zero the residual spectrum obtained after all the
identified lines have been subtracted (Figure~\ref{continuum}, Panel (c)). The
area of a line measured with respect to the effective continuum
(Figure~\ref{continuum}, Panel (d)) should be a more valid estimate of the correct EW of the particular
atomic transition we are interested in, once the uncertainties of the procedure
are properly taken into account. 

\subsubsection{Experiment with Artificial Spectra}
\label{sec-arti}

\begin{figure}[t]
\includegraphics[angle=-90,scale=0.30]{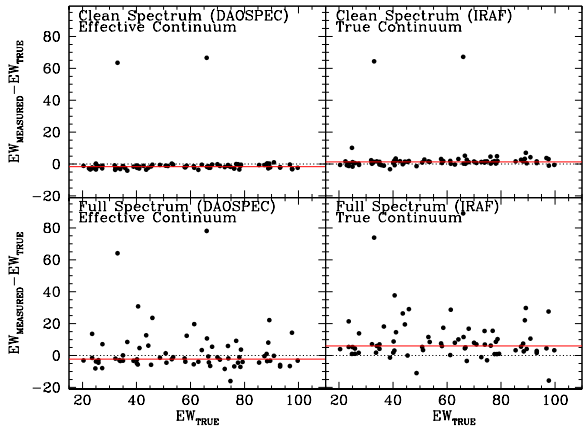}
\caption{Difference between the measured and true (input) EW on the artificial
spectra. The left panels show DAOSPEC measurements, based on the {\em effective
continuum}, and the right panels IRAF {\em splot} measurements based on the {\em
true continuum}. Top panels show measurements on the {\em clean spectrum} and
bottom panels show measurements on the {\em full spectrum}. In each Panel, the
dotted lines mark a perfect agreement ($\Delta$EW=0), while the solid lines mark
the median $\Delta$EW. \label{ew-cont}}
\end{figure}

We tested the above assertions with a simple experiment employing artificial
spectra, created using a list of more than 1000 atomic lines belonging to 50
species, in the wavelength range 5500--5600\AA. The atomic data were obtained
from the VALD\footnote{{\tt http://www.astro.uu.se/$\sim$vald/}} database
\citep{vald}. The most recent version of the spectral synthesis code originally
described in \citet{spite} was employed, together with OSMARCS model atmospheres
\citep[][and following updates]{marcs}, to predict the EW of each line for a
metal-rich giant that should represent the typical difficult case:
$T_{\rm{eff}}$=4500~K, $\log g$$=$1.0, $v_t$$=$1.0~km~s$^{-1}$, and [M/H]=0.0,
with no $\alpha$-enhancement. The artificial spectra were generated as simply as
possible, simulating lines with EW$<$100~m\AA\  with saturated Gaussians
(Section~\ref{dao-ew}), and chopping out stronger lines (discussed in
Section~\ref{dao-ew}). A flat true continuum level was set to 1.0. The simulated
spectral quality is typical of a modern echelle spectrograph, FWHM=0.15~\AA\ 
($R\simeq40,000$), S/N=35, a scale of  0.03~\AA/pixel. Two different spectra
were generated (Figure~\ref{arti}): a {\em clean spectrum}, where all lines
smaller than 20~m\AA\  were omitted, and a {\em full spectrum} including
all lines, even the hundreds of tiny lines measuring as little as a fraction of a m\AA. 

The same list of 86 lines was then measured on both the {\em clean spectrum} and
the {\em full spectrum}: once with DAOSPEC relative to its estimate of the
effective continuum, and again with the task {\em splot} in IRAF relative to the
true continuum, which is exactly known and fixed to 1.0 in this ideal case (see
Figure~\ref{arti}). Results are presented in Figure~\ref{ew-cont}. As
anticipated, DAOSPEC's effective continuum is lower than IRAF's---by 1\% in the
clean spectrum and by 2.5\% in the full spectrum. Both differences are smaller
than the corresponding root-mean-square residuals estimated by DAOSPEC, which
are 2.4\% per pixel for the clean spectrum and 2.9\% per pixel for the full
spectrum. Since the simulated noise was the same in both spectra, the slightly
higher variance measured in the full spectrum is presumably due to the
hundreds of additional small, unresolved lines that it contained.

Both methods of measurement gave similar results on the {\em clean spectrum}
(top panels of Figure~\ref{ew-cont}): lines measured with respect to the
effective continuum by DAOSPEC were underestimated by
$\Delta$EW$=$--1.60$\pm$0.95~m\AA\  while lines measured with respect to the
true continuum by IRAF/{\it splot\/} were overestimated by
$\Delta$EW$=$1.45$\pm$0.95~m\AA. Both differences are probably negligible
compared to the uncertainties involved ($\sim 1.5\sigma$).  


When considering the {\em full spectrum} however, apart from an overall increase
in the spread of the measurements, we find that the DAOSPEC measurements
relative to the effective continuum remain closer to the input values and more
consistent with the measurements on the {\em clean spectrum} than the IRAF/{\it
splot\/} measurements with respect to the known true continuum.  Specifically,
we find net shifts of $\left<\Delta\hbox{\rm EW}\right> =-2.3\pm5$~m\AA\ for
DAOSPEC/effective continuum and $\left<\Delta\hbox{\rm EW}\right> =
+6.0\pm5.7$~m\AA\ for splot/true continuum. We assert, therefore, that as
crowding increases, measurements relative to the true continuum---even when it
is known perfectly---become unreliable because of the likelihood of including
opacity contributions from unrecognized and unrelated spectral lines in the EW
measurement for the target line\footnote{In reality, since perfect knowledge of
the true continuum position in real---as opposed to simulated---spectra is
impossible, a person who attempts to estimate the true continuum level to anchor
EW measurements risks a systematic error that is probably somewhere between the
two limiting cases shown here.}.  

This is especially important to consider when one is measuring EW interactively
with packages such as IRAF, MIDAS or SPECTRE. The procedure of readjusting the
continuum locally around the line can produce a systematic bias if the user
tries to ignore weak spectral lines in the continuum samples by favoring the
highest points in the spectrum or by selective placement of the continuum-sample
endpoints; this fails to allow for the possibility that flux may be removed by similar
uninteresting but nevertheless real absorption features occupying the same
wavelength range as the line of interest.  The depression of the continuum
by such weak lines is, in fact, very difficult to estimate in an impersonal and
repeatable way by eye when using only small regions around each line.  However,
we believe that we can reduce the systematic errors due to unrecognized
absorption features by iteratively using our model profiles to subtract all
recognized spectral lines and then running the estimated continuum through the
middle of the remaining flux spectrum, taking this to represent the typical
local flux per pixel in the presence of all {\em unmodeled} opacity sources.

\subsection{EW Gaussian Fit}
\label{dao-ew}

When measuring EWs, DAOSPEC applies to crowded stellar spectra a concept also
used in the analysis of crowded stellar images \citep[e.g., DAOPHOT and
ALLSTAR,][]{daophot}, i.e., the idea of fitting small groups of features
simultaneously. This concept is used also in the interactive deblending routine
of the IRAF task {\em splot}, but to our knowledge has not been implemented in
an automatic program up to now. Besides allowing each absorption feature to
be responsible for its own fraction of the flux missing from individual pixels,
another advantage of this approach is that the fitting function of each feature
now does not need to be truncated close to the feature center, but
can extend outwards by several pixels (e.g., up to 6$\sigma$ for a Gaussian
function, where the intensity of a line is down by a factor $e^{-18} \sim
10^{-8}$ from its maximum value), even in those regions where it overlaps one or
more nearby features. Deblending as in DAOSPEC can therefore be more effective than
in those programs that fit lines individually.

The fitting function employed in DAOSPEC is based on a {\em saturated
Gaussian}\footnote{In the future, when higher resolution spectrographs will
become routinely available, we plan to update DAOSPEC so that the user will be
able to choose among different fitting profiles, including the Voigt profile,
that would take care of lines with extended wings.  It should be noted that these
extended wings will substantially increase the analysis times which may, we
hope, be ameliorated by faster computers.} defined as

\begin{equation}
h(\lambda)=\frac{g(\lambda)}{1+g(\lambda)}
\end{equation}

\noindent  
where $g(\lambda)$ is the Gaussian function (Equation~\ref{Gauss}). This
particular formulation was chosen because deviations of $h(\lambda)$ from the
original Gaussian shape are only important for strong lines: accordingly, as
$g(\lambda)$ becomes small $h(\lambda)$ tends to $g(\lambda)$.  Futhermore, this formula
ensures that the residual flux at line center ($\sim 1-h(\lambda)$; cf. Eq.~\ref{Gauss2})
approaches zero asymptotically as the EW grows without limit; if a pure Gaussian
function had been retained as the model spectral line, there is no mathematical
principle that prohibits the line profile from crossing over into negative
fluxes ($A > 1$ in Eq.~\ref{Gauss}) as the line strength increases.  Clearly,
our adopted numerical formulation is not justified from physical first
principles, but it is found empirically that it provides a useful extension
of the general method to moderately saturated lines.  It cannot, of course,
adequately reproduce lines that are strongly saturated. The exact EW at which
the method is no longer valid clearly depends on the spectral resolution and S/N
ratio, since the better a spectral line is defined, the more statistically
significant its deviation from the model profile will be\footnote{Of course, not all
lines deviate perceptibly from the adopted profile at exactly the same EW, since
the line profile, and in particular the extended line wings, depend on the
physical properties of the particular transition that produces the line.}.

\begin{figure}[!t]
\includegraphics[scale=0.35]{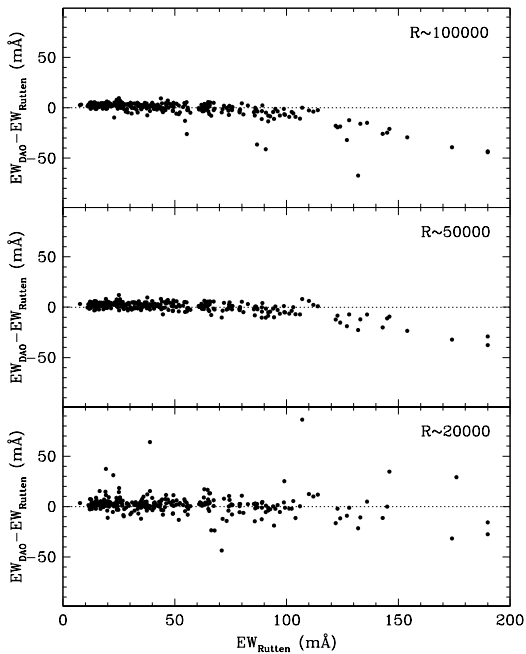}
\caption{Comparison between DAOSPEC measurements and EWs by
\citet{rutten} on the UVES Solar spectrum for three different
resolutions: R$\sim$100000 (top panel), R$\sim$50000 (middle panel)
and R$\sim$20000 (bottom panel). The EW at which the Gaussian
approximation adopted by DAOSPEC is no longer valid appears clearly as
that EW where $\Delta$EW starts to deviate from zero.
\label{fig-Gauss}}
\end{figure}

To test how well our Gaussian approximation holds for real spectra, we have used
the UVES Solar spectrum\footnote{\tt
http://www.eso.org/observing/dfo/quality/UVES/pipel ine/Solar\_spectrum.html},
which has an intrinsic resolution of about R$\sim$10${}^5$; we have also degraded
it to resolutions of R$\sim$50000 and R$\sim$20000. We have measured EW with
DAOSPEC using the list of clean Solar lines by \citet{rutten}. The difference
between measurements by DAOSPEC and by \citet{rutten} is shown in
Figure~\ref{fig-Gauss} for the three Solar spectra. As can be seen, for the two
high resolution cases DAOSPEC measurements begin to deviate from those of
\citet{rutten} at around 80--90~m\AA\  (for R$\sim$10${}^5$) and 100--120~m\AA\
(for R$\sim$50000), while the agreement remains within the noise up to
$\sim$200~m\AA\ in the lowest resolution case. 

Another feature of DAOSPEC that helps deblending in crowded spectra is the fact
that the FWHM is consistent for all lines, instead of being left as a free
parameter in the fit of each individual line, as was already discussed in Section~\ref{int-find}. 
Either the FWHM is forced to be the same for all lines, or it can scale with
wavelength, which is more appropriate for echelle high-resolution spectra that
have been rebinned in wavelength. A first guess at the FWHM is input by the user
and it is subsequently refined  during the robust least-squares fits that
take place in the main loop of DAOSPEC (central part of Figure~\ref{flow}).

An additional advantage of the way DAOSPEC works again relates to the
measurement of strong features in crowded spectra.  For interactive routines
trying to fit each absorption line separately, the safe region is generally
limited to the line core, which is most often approximately Gaussian. This leads
to an underestimate of the EW of strong lines that have an important flux
contribution in their (non-Gaussian) wings. DAOSPEC fits lines by minimizing the
residuals over the entire line profile after all fitted lines have been subtracted:  
the {\it total\/}
area of the residuals is forced to be as close to zero as possible (an oversubtracted
core forced to balance undersubtracted wings), and the total EW
is still approximately correct to first order even if the line profile deviates
from the adopted shape. This results from the fact that the whole line, including
its extended wings, is included in the fitting region. Of course, as already noted, very
strong lines cannot be properly measured this way.

\subsection{Uncertainties}
\label{dao-err}

DAOSPEC provides three pieces of information that can be used to characterize the
uncertainty of each EW measurement, described in the Sections below. 

\subsubsection{Formal Standard Error}

\begin{figure}[t]
\includegraphics[angle=0,scale=0.35]{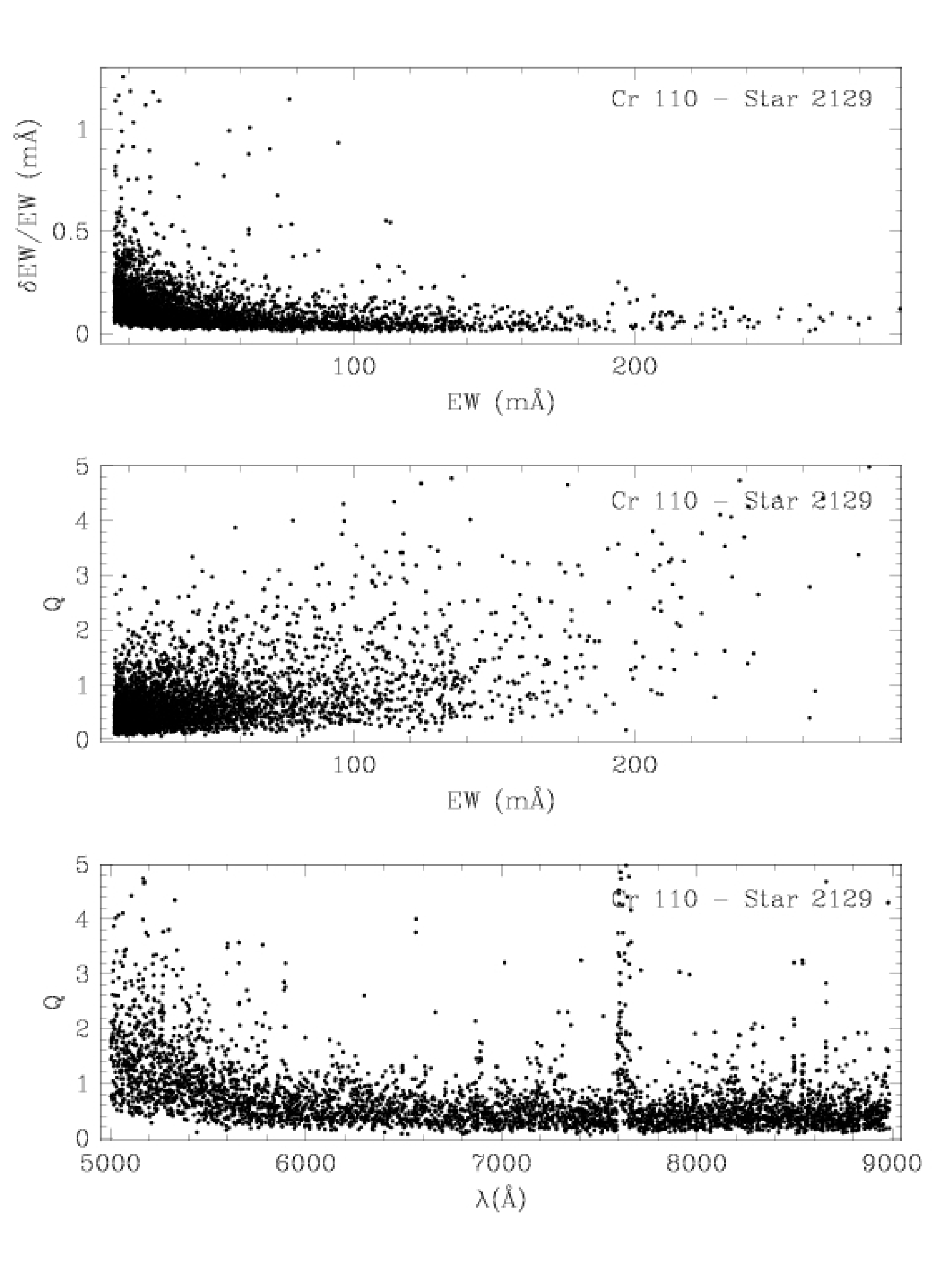}
\caption{Behaviour of the relative error $\Delta$EW/EW as a function
of EW (top panel) and of the quality parameter $Q$ as a function of
EW (middle panel) and $\lambda$ (bottom panel). The plot reports
DAOSPEC measurements for star 2129 in the Open Cluster Cr~110 (from
Pancino et al., in preparation). \label{quality}}
\end{figure}

The first is the formal standard error, $\sigma$(EW), obtained during the
least-squares determination of the EW of each line. This is a purely empirical
$\sim 68$\% confidence interval on the derived value of the line's EW, derived
from the standard deviation of the local flux residuals remaining after the
fitted continuum and all detected lines have been removed from the  observed
spectrum. Like EW, this has units of wavelength and results from a standard
least-squares propagation of errors that weights the residual of each individual
pixel according to that pixel's contribution to the determination of the line's
EW. The relative error $\sigma$(EW)/EW can be used as an effective means of
distinguishing good and bad lines, as shown in the top panel of
Figure~\ref{quality}. Weaker lines have increasingly large relative errors, as
expected due to the presence of noise and random blending.  When the error rises
above, say, 20\% the lines could be considered unsatisfactory and could be
omitted from an abundance  analysis\footnote{As a rule of thumb, lines with EW
smaller than three times $\sigma$(EW) should certainly be rejected.}. Another
way of using the relative error given by DAOSPEC would be to derive the
abundance of one particular element not as a straight average of the single-line
abundances, but as a weighted average, using $\sigma ({\rm EW}) d({\hbox{\rm
abundance}})/d({\hbox{\rm EW}})$ to define the weights.  

\subsubsection{Quality Parameter}

The second is called the {\em quality parameter} $Q$, and is also associated
with each individual EW measurement. For each line, the root-mean-square value
of the intensity residuals is calculated for the same range of pixels as were
employed in the profile fit for that line. The ratio of this root-mean-square
residual to the root-mean-square intensity residual for the spectrum as a whole
is that line's $Q$ value. Thus, it tells whether the particular region where
that line sits is more ($Q>1$) or less ($Q<1$) problematic than the average for
the spectrum as a whole.

The $Q$ value for a given line can be higher than average if the line more
strongly deviates from the Gaussian form than the average line, because in this
case the local residuals will have a higher variance than the rest of the
spectrum.  This is illustrated in the middle panel of Figure~\ref{quality},
where the $Q$ values of lines measured in star 2129 of the Open Cluster Cr~110
(from Pancino et., in preparation) are plotted as a function of EW. As can be
seen, most lines above 200~m\AA\ have $Q>1$, that is, the residuals near their
positions are higher than the average residuals for other parts of the spectrum. 

Another reason why $Q$ could be higher than average is that the spectral quality
(crowding, S/N ratio, defects) is not homogeneous along the whole wavelength interval.
In the bottom panel of Figure~\ref{quality}, $Q$ is plotted as a function of
wavelength. As can be seen, lines at the blue end of the spectrum---where the S/N
ratio is lower than in the rest of the spectrum---tend
to have higher $Q$ values than the rest.  Also,
the region around 7600~\AA\ tends to have bad $Q$ values due to the telluric
absorption band of O$_2$. In some cases (i.e., strongly varying S/N ratios) these
effects could be alleviated by adopting different $Q$ selections for different
spectral regions. In some other cases (anomalous crowding, defects or molecular
bands), it could be safer to cut out the whole region in $\lambda$ where
anomalous $Q$ values are clustered.  

As a general rule, therefore, a lintel in $Q$ can be defined, depending on the
particular spectrum, to reject less reliable measurements. In the case of
Figure~\ref{quality}, for instance, some value 1$\ltsim$Q$_{\hbox{\small
lim}}\ltsim$2 could be adopted depending on the specific goals of the
abundance analysis.  However, please note that if a too strict selection in
$Q$ is adopted, some systematic effects could be introduced in the analysis
that are difficult to foresee. Perhaps safer would be simply to increase the
adopted $\sigma$(EW) values for lines with moderately large values of $Q$ in
the ensuing abundance determinations.  

\subsubsection{Standard deviation of Residuals}
\label{sec-conterr}

The third parameter, which characterizes the quality of the solution as a whole
and is given in the header of the output file, it is the standard deviation of
the flux residuals, which can be constructed as the r.m.s.\ relative
(percentage) scatter in the effective continuum on a pixel-by-pixel basis. This
has units of flux, and is mainly due to photon noise in the spectrum, of course,
but substantial contributions can also come from unrecognized weak lines and
molecular bands, which are generally more important in metal-rich, cool, and/or
low-gravity stars, or from imperfections of the instrument or in the preliminary
analysis, such as in the  merging of echelle orders, the subtraction of
night-sky emission lines, or the removal of CCD fringing effects. Note that
since this index represents a pixel-by-pixel residual variance about the adopted
mean continuum, positive and negative EW errors should occur with similar
frequency, and the overall error in the abundance of any given atomic species,
due to these effects, should decline with increasing numbers of lines measured,
expecially if the lines span a large wavelength interval. 

To evaluate the likely impact on the EW results due to the overall, {\em
systematic} uncertainty of the continuum placement, a few considerations can be
useful. First of all, the {\em absolute} uncertainty of the continuum level, in
flux units, must be at least as large as the standard deviation of the flux
residuals divided by the square root of the number of pixels:

\begin{displaymath}
\sigma(I_C) \geq {\sigma({\hbox{\rm 1 pixel}})\over \sqrt N}
\end{displaymath}

\noindent 
Since it is often more convenient to work in relative units, with
$\sigma({\hbox{\rm 1 pixel}})/I_C$ representing the {\em fractional} standard
deviation of the flux residuals,

\begin{displaymath}
{\sigma(I_C)\over I_C} \geq {{\sigma({\hbox{\rm 1 pixel}})/I_C} \over \sqrt N}
\end{displaymath}

\noindent
gives a lower limit to the {\it relative} uncertainty of the continuum
placement.

However, the $N$ pixels have been used to derive a model continuum consisting of
a polynomial of $n^{th}$ order, i.e., to determine $n+1$ free parameters. To be
ultraconservative, the pixels that are covered by spectral lines could also be
entirely omitted from the uncertainty computation. Therefore, if $n_l$ is the
total number of lines found by DAOSPEC and $n$ is the polynomial order, the
ultraconservative estimate of the uncertainty of the global continuum placement
procedure is

\begin{equation}
\frac{\sigma(I_C)}{I_C} \ltsim \frac{\sigma(\hbox{1~pixel})/I_C}{\sqrt{\frac{N-(n_l\cdot FWHM)}{n+1}}}.
\end{equation}

where we have supposed that each of the $n+1$ parameters of the continuum model
is independently determined from at most $N\over{n+1}$ pixels. For instance,
in a modern spectrum containing 40,000 pixels ($N$) and 1,000 spectral lines
($n_l$), having a FWHM of 10~pixels and a continuum shape corresponding to a
$20^{th}$-order polynomial, and producing a residual spectrum with a noise level
of 1\% (per pixel, root-mean-square), the overall uncertainty in the continuum
placement cannot be as good as 1\%/$\sqrt{40000} \approx 0.005$\%, but neither
is it likely to be as bad as 1\%/$\sqrt{{40000 - (1000)(10)}\over{21}} \sim
0.03$\%. The global impact of this continuum uncertainty on the EW results can
be estimated by comparing a relative standard error calculated in this way  to
the typical biases illustrated in Fig.~2 above. Alternatively, the spectrum,
normalized with the continuum produced by DAOSPEC, can be multiplied by the
corresponding $1-\Delta I_C/I_C$ and by $1+\Delta I_C/I_C$ and the two spectra
thus obtained can be fed to DAOSPEC while switching off the continuum-fitting
procedure (ORDER=--1). The difference between the two sets of EW and the
original DAOSPEC result will give a more precise evaluation of the systematic
uncertainty in the EWs due to the continuum placement procedure.  

\begin{figure}[t]
\includegraphics[angle=270,scale=0.30]{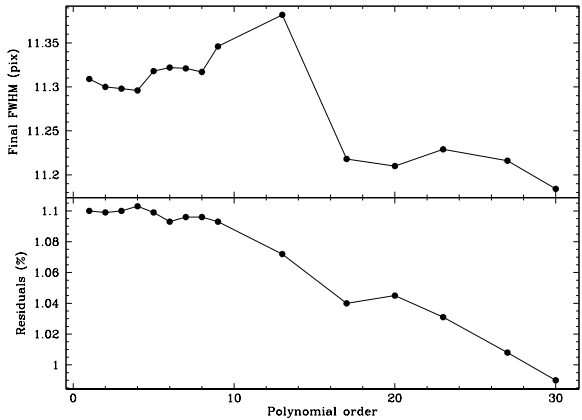}
\caption{An example that shows how the final output FWHM varies with the
Legendre polynomial order (top Panel) and how the residuals of the subtracted
spectrum (bottom Panel) vary with the Legendre polynomial order. \label{order}}
\end{figure}

\begin{figure}[t]
\includegraphics[angle=270,scale=0.30]{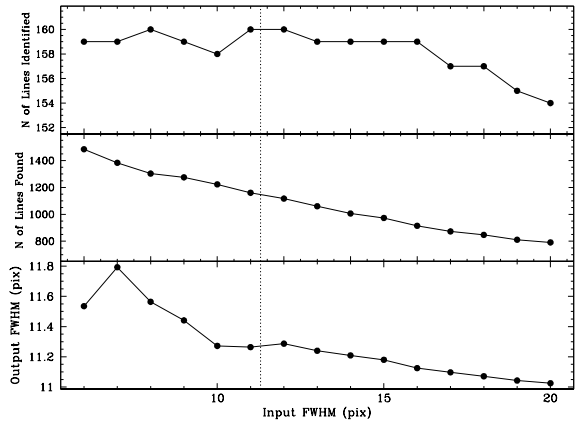}
\caption{The three panels show (from top to bottom) how the number of lines
identified in the input line list varies with the input FWHM, how the total number
of lines found varies with the input FWHM and how the final output FWHM varies
with the input FWHM. The dotted lines mark the correct FWHM=11.3 pixels. 
\label{fwhm1}}
\end{figure}

\subsection{Dependency on Input Parameters}
\label{dao-pars}

In this section we perform a few experiments to show how the results of DAOSPEC
depend on the most relevant input parameters. A detailed description of the
parameters can be found in the DAOSPEC Cookbook \citep{cookbook}, here we 
describe only the ones that have the most impact on the EW results.

\subsubsection{Polynomial Order for Continuum Fit}
\label{dao-cont}

As already mentioned in Sections~\ref{sec-ew}, \ref{placement} and
\ref{sec-con}, overall continuum placement has a global and systematic impact 
on the EW measurements. The only user-specified DAOSPEC input parameter that
directly regulates it is the order of the Legendre
polynomial used. This has little impact on the {\em average} continuum level
because there are no other rejection or tweak parameters that alter the fit (see
Section~\ref{sec-dao} for a description on how the continuum is fit). This is
illustrated in Figure~\ref{order}, where a sample spectrum has been fitted with
different-order polynomials and the effect on the output FWHM and residuals of
the subtracted spectrum has been shown. In general the FWHM changes are erratic,
while the average residuals decrease when the polynomial order increases.

To choose the appropriate order for the continuum fit, a good rule of thumb is
to compare the scale length of the global variations in the spectral shape 
to the total spectrum length. If the variations are frequent and
happen down to a scale length of $1/m$ of the total length of the spectrum,
then the appropriate order to choose is $m$. If too low a value of $m$ is adopted,
there will be regions of the spectrum where the continuum is overestimated and
other regions where it will be underestimated, leading to an increase in the
scatter of the resulting EWs (see also Section~\ref{sec-sou}).
For chemical species whose abundances rely on only a few lines, the mistake can
be significant. On the other hand, provided that the spectrum contains many,
many more than $m$ valid pixels (which should normally be the case) the main
risk in adopting too {\it large\/} a value of $m$ would be some increase in the
reduction time.  In case of doubt, we recommend that the optional output
spectrum containing the fitted continuum be produced and overplotted on the
original spectrum with, e.g., IRAF to check whether the order has been chosen
appropriately (see also Figure~\ref{fig-feros}).  

\begin{figure}[t]
\includegraphics[angle=270,scale=0.30]{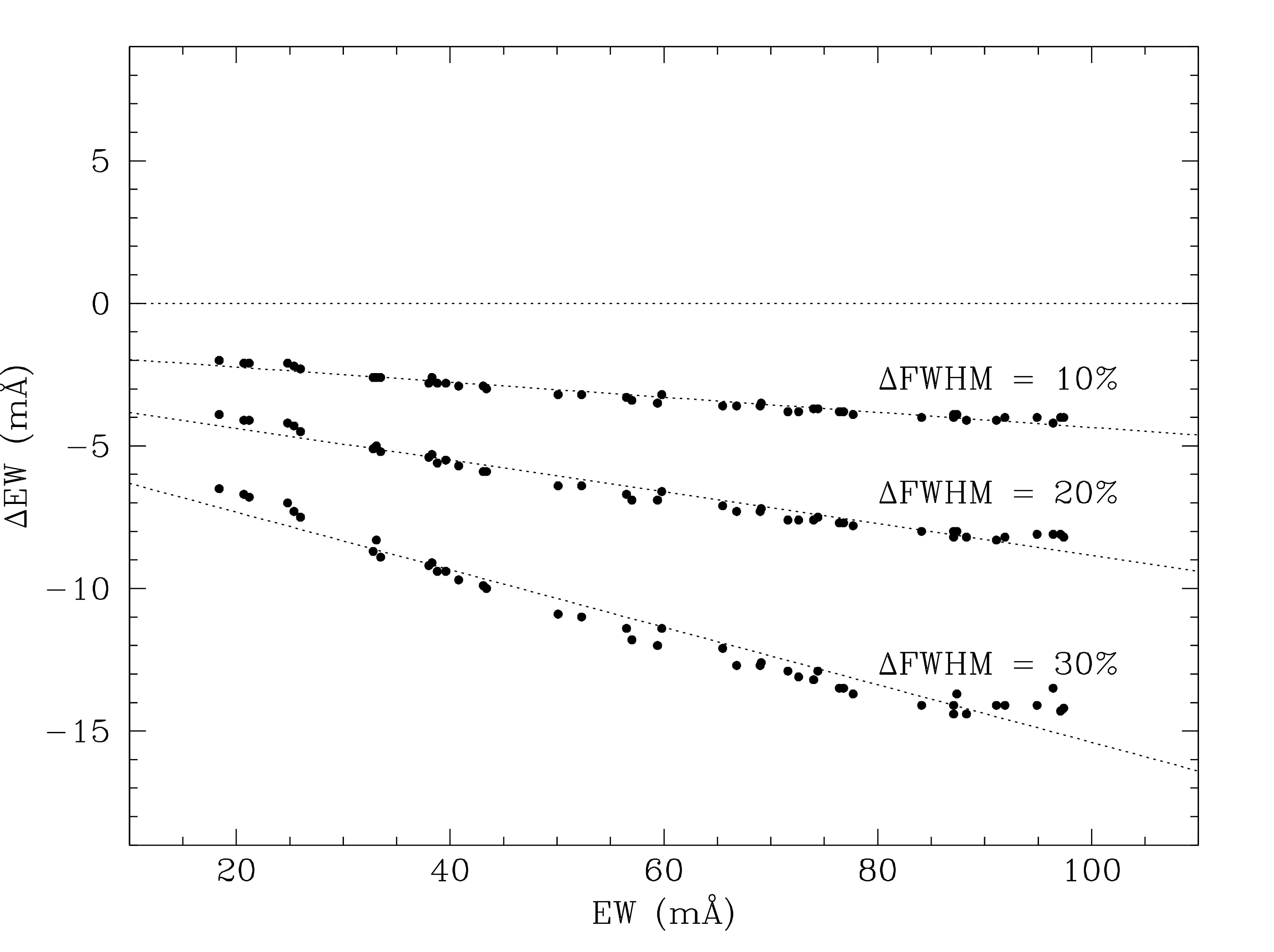}
\caption{A spectrum with a FWHM$\simeq$5 pixels is measured with DAOSPEC using a
FWHM which is underestimated by 10\%, 20\% and 30\%. \label{fwhm2}}
\end{figure}

\subsubsection{First-Guess FWHM and Scaling}
\label{sec-fwhm}

DAOSPEC needs a first guess for the FWHM (in pixels) to be able to initially
distinguish and ignore features that are either too narrow or too broad to be
likely real astrophysical spectral lines (see Section~\ref{sec-dao}). 
Figure~\ref{fwhm1} shows what happens when the input FWHM is wrongly estimated
by a large factor: while the number of detected lines that also appear in the
input line list does not change very much, there is a significant change in the
total number of lines found by the code; presumably this is because the input
list contains primarily ``interesting"---i.e., relatively strong---features, and
it is the discovery of weak, mostly anonymous lines that is sensitive to the
assumed FWHM.  Since the working line list is limited to those lines detected
initially, non-detection of real lines caused by initially specifying too large a FWHM is
not fixed during the program's main iteration loop.  However, spurious
detections occasioned by the initial assumption of too small a value of the FWHM
may be rejected later on, as the FWHM and the EWs are refined.  
We find that the {\em output} FWHM does not change significantly if
the {\em input} FWHM is wrong by one pixel (in this case, $\sim$10\%), but
starts to be over/underestimated perceptibly when the input FWHM is wrongly
estimated by more than that. Comparing the {\em input} and the {\em output} FWHM
after a first run of the code can help in deciding whether the reduction should
be reperformed with a different input FWHM.  

The characteristic FWHM of a spectrum is often considered of secondary
importance: we are not usually interested in its value except as it relates to
the quality of the EWs that are the principal goal of the analysis. However,
DAOSPEC could be used to estimate the FWHM of a spectrum, which is extremely
useful for spectral synthesis. While we have seen (Section~\ref{sec-ew}) that a
wrong {\em average} level of the continuum produces a {\em systematic offset} of
the EW that does not depend strongly on EW, we show here that a wrong fitted
value of the FWHM produces a {\em systematic trend} in the EW that depends on
the EW itself. Figure~\ref{fwhm2} illustrates the difference in the EW on lines
measured on the {\em clean spectrum} described in Section~\ref{sec-arti} and
Figure~\ref{arti}, when the input FWHM {\em is kept fixed to a wrong value},
i.e., underestimated by 10\%, 20\% and 30\%\footnote{Please note that in this
case, unlike in Figure~\ref{fwhm1}, both the {\em input} and {\em output} FWHM
are wrong, i.e., we force DAOSPEC to use the wrong values by setting FIX=1}.
Therefore, if the FWHM to which DAOSPEC converges differs by more than
approximately 10\% from the user input FWHM, the measurement should be repeated
until the input and output FWHM are sufficiently similar.  

\begin{figure}[t]
\includegraphics[scale=0.37]{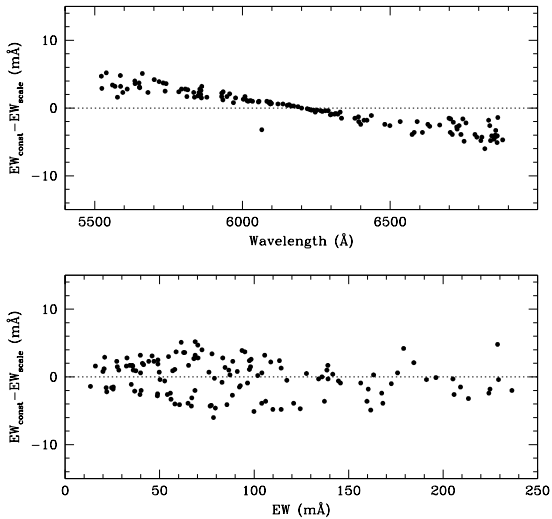}
\caption{EW difference on measurements done with constant FWHM and with FWHM
scaled with $\lambda$ on an echelle spectrum of Star 2129 in Cr~110. In the top
panel $\Delta$EW is plotted versus wavelength, showing how EWs are overestimated
in the blue and underestimated in the red. In the bottom panel $\Delta$EW is
plotted versus EW, where the mistake appears as an extremely large scatter.
\label{scale}}
\end{figure}

One other important point to keep in mind is that in an echelle spectrum, where
$\lambda / \delta \lambda$ is constant and the pixels are rebinned linear in
wavelength, the FWHM changes from the blue to the red side of the spectrum. DAOSPEC
allows the use of a constant FWHM or of a FWHM that scales with wavelength.
Choosing a constant FWHM on an echelle spectrum would produce an overestimate of
the FWHM in the blue and an underestimate on the red, with only the central EWs
measured properly. This is illustrated in Figure~\ref{scale}, where the echelle
spectrum of star 2129 in Cr~110 (Pancino et al., in preparation), ranging from 5500
to 6900~\AA, is measured without scaling the FWHM and compared with the correct
measurements where FWHM is scaled. As can be seen, the longer the spectrum, the
bigger the mistake, which can be of a few m\AA. When the EW differences are plotted
as a function of wavelength (top panel), we see a clear trend, with an increase of
the spread on the sides, while when they are plotted as a function of EW (bottom
panel) the only noticeable effect is an enormous increase in the spread.

\begin{figure}[t]
\includegraphics[angle=270,scale=0.30]{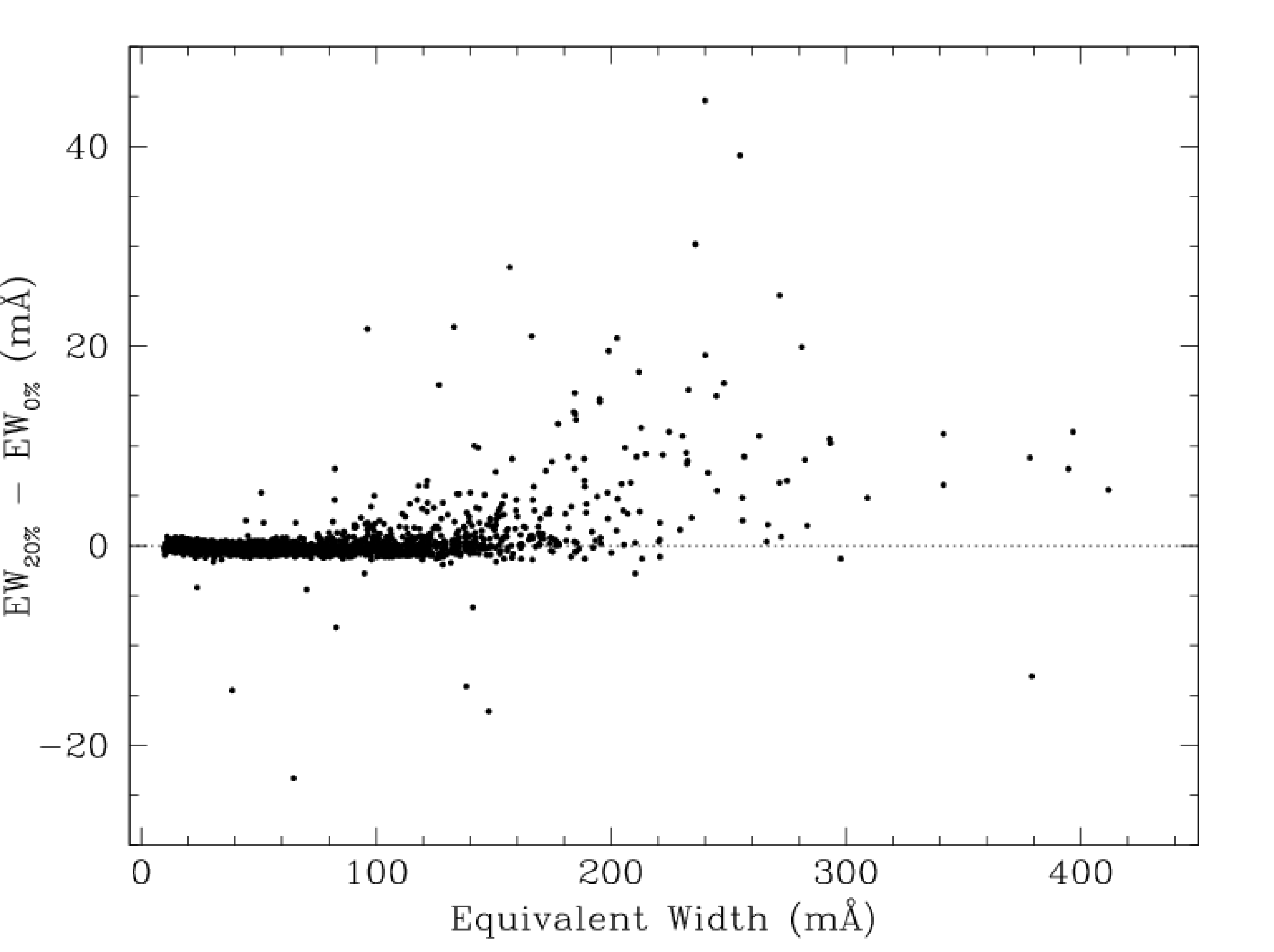}
\caption{EW difference on measurements done with the correct Residual Core
Flux value of 20\% and the wrong value of 0\%. The largest differences appear of
course for strong lines.\label{core}}
\end{figure}

\subsubsection{Residual Core Flux}

In  real stellar spectra, the center of a line is never completely black because
the effective surface of a stellar atmosphere is not at a temperature of
absolute zero. Even strongly saturated lines, such as the H$_{\alpha}$ line (at
6563~\AA) in the spectrum of a low mass red giant, never reach zero flux even in
their cores.  Telluric absorption lines (i.e., the O$_2$ and H$_2$O bands) or
interstellar absorption lines (e.g., the NaD doublet) can, on the other hand, be
much darker in their centers. Depending on the resolution and pixel size of a
particular spectrum, the maximum depth of strong lines varies a bit. For
example, the spectrum of Figure~\ref{core} has an H$_{\alpha}$ line that reaches
20\% of the continuum value in the line core.  This effect is modeled in the
DAOSPEC analysis by a further simple modification of Eq.~(6) above:  the model line
profile now becomes

\begin{equation}
f(\lambda)=\frac{g(\lambda)}{1+g(\lambda)/(1-\rho)}
\end{equation}

The model of the spectrum in the region of the line then becomes

$$I(\lambda) = I_C \cdot (1 - f(\lambda)) \rightarrow \rho ~ I_C$$

\noindent as $g(\lambda) \rightarrow \infty$.  The quantity $\rho$, known as the
``residual core flux,'' is specified by the user and is typically in the range
5--25\%.  It ensures that the profiles of strong lines saturate to a value
equal to some fraction of the local continuum flux $I_C$ rather than to a value
of zero.  A side-effect of this saturation model is that as lines become stronger
within the saturation regime, they also become broader even for a fixed value of the
FWHM parameter.  This improves to some extent the capability of the model line profiles
to mimic the actual profiles of real spectral features.

\subsection{Dependency on Spectral Quality}
\label{dao-qual}

To test the effect of the spectral quality on the EW measured by DAOSPEC, we
created different versions of the {\em full spectrum}, our most crowded artificial
spectrum described in Section~\ref{sec-arti}, with different S/N ratios,
resolution (R=$\lambda / \delta \lambda$), and pixel sampling. 

\begin{figure}[t]
\includegraphics[scale=0.35]{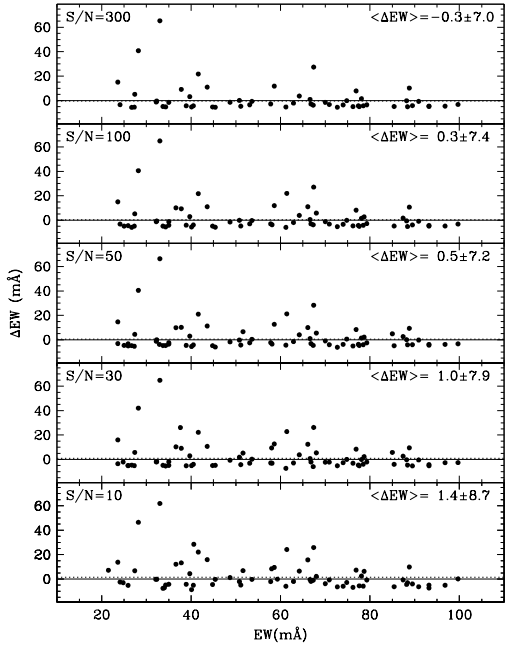}
\caption{Difference on the EW measured with DAOSPEC and the ``true" input EW of
five artificial spectra with R$=35000$ and S/N ratios ranging from S/N$=300$ (top panel)
to S/N$=$10 (bottom panel). Continuous lines mark perfect agreement and dotted
lines the $\sigma$ clipped average, also indicated on the top of each panel.
\label{daosn}}
\end{figure}

\subsubsection{S/N Ratio}
\label{sec-sn}

The first experiment is shown in Figure~\ref{daosn}, where the difference between
the EW measured by DAOSPEC and the {\em known} input EW is plotted versus EW for
five artificial $R\simeq35000$ spectra at various S/N ratios and with a sampling
of approximately 5 pixels for each resolution element. Mean ($\sigma$-clipped)
values with their line-by-line variance ($\sigma$) are reported in
Figure~\ref{daosn}. As can be seen, there is no significant variation of the
average $\Delta$EW, which always remains close to zero, within the uncertainties.
The variance $\sigma$ increases by approximately 2~m\AA\  from the S/N$=$300 to
the S/N$=$10 case, as expected.  Therefore, DAOSPEC gives good results even with
a S/N ratio as low as 10. This is due in part to the fact that the FWHM used by DAOSPEC is
the same\footnote{Scaled with wavelength in the case of echelle spectra.} for all
lines: if the spectra are properly sampled\footnote{at least 2 pixels per
resolution element.}, DAOSPEC can usually tell a line from a noise feature. This
result is somewhat in contradiction with a similar experiment performed by
\citet{sousa}, and described in detail in Section~\ref{sec-sou}. 

\begin{figure}[t]
\includegraphics[scale=0.35]{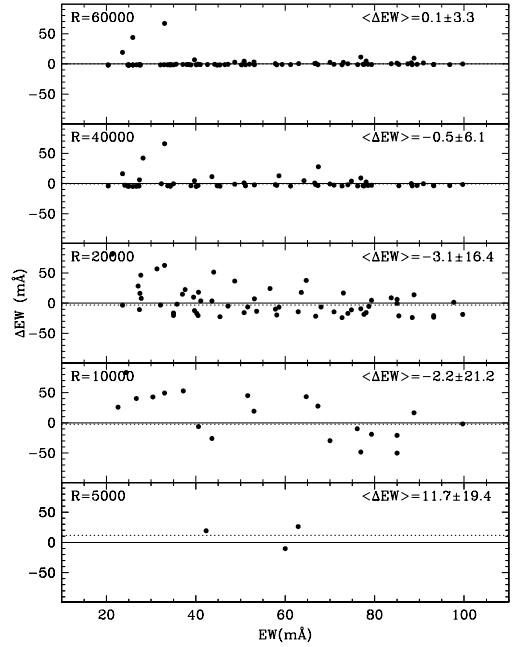}
\caption{Difference on the EW measured with DAOSPEC and the ``true" input EW of
five artificial spectra with S/N$=$100 and resolution ranging from R$=$60000 (top
panel) to R$=$5000 (bottom panel). Continuous lines mark perfect agreement and
dotted lines the $\sigma$-clipped average, also indicated on the top of each
panel. \label{daores}}
\end{figure}

\subsubsection{Resolution}
\label{sec-res}

Resolution appears to be more critical than the S/N ratio as can be seen in 
Figure~\ref{daores}, where five spectra of fixed S/N$=$100 and similar sampling
of approximately 5 pixels have been created at R ranging from 5000 to 60000.
Again, the average $\sigma$-clipped difference between DAOSPEC measurements and
the ``true" input EWs are always consistent with zero within the spread.
However, for R$=$5000, the total number of identified lines is very low. This,
of course, is inherent to the intrinsic quality of the spectra and not
necessarily specific to the algorithms employed in DAOSPEC. Nevertheless,
R$=$10000 can perhaps be considered a lower limit of validity of the EW method
in general, and of the use of DAOSPEC in particular.  We expect that stars with
very high rotation velocities will behave as if they had a lower instrumental
resolution, as a consequence of the correspondingly broad line profiles. 

When resolution reaches R$\simeq$40000, $\Delta$EW gets definitely better not
only in terms of the average, but also in terms of the spread, which at
R$=$60000 becomes $\simeq$3~m\AA. However, we must reiterate that our artificial
spectra are simplistic, because all the simulated lines have pure Gaussian
profiles. In real spectra, the effect of non-Gaussian wings on strong lines
becomes important, as has been discussed in Section~\ref{dao-ew} and shown in
Figure~\ref{fig-Gauss}. 

\begin{figure}[t]
\includegraphics[scale=0.35]{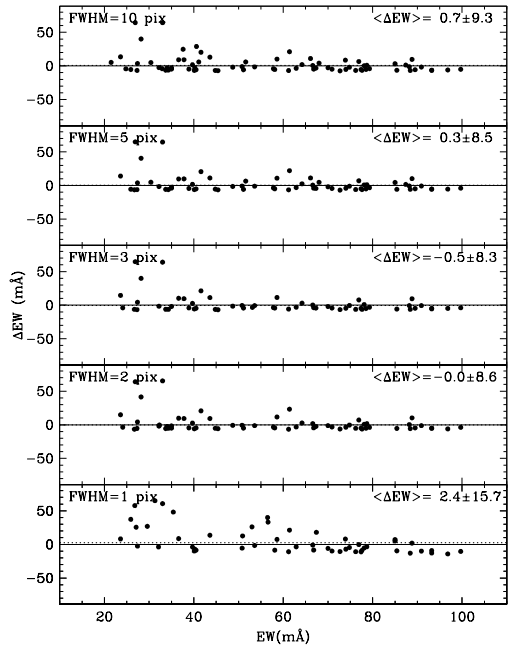}
\caption{Difference on the EW measured with DAOSPEC and the ``true" input EW of
five artificial spectra with S/N$=$100, R$=$35000 and pixel sampling ranging from
FWHM$=$10~pix (top panel) to FWHM$=$1~pix (bottom panel). Continuous lines mark
perfect agreement and dotted lines the $\sigma$ clipped average, also indicated on
the top of each panel. \label{daopix}}
\end{figure}

\subsubsection{Pixel Sampling}
\label{sec-pix}

The last factor that may have an impact on EW measurements is the pixel sampling
of the resolution element. The Nyquist-Shannon Theorem \citep{shannon} states
that the reconstruction of a continuous signal from its samples is possible if
the signal is bandlimited and the sampling is greater than twice the signal
bandwidth.  In our case, the adequate reconstruction of a line profile is only
possible if it is sampled by at least two pixels. 

Figure~\ref{daopix} shows the usual comparison between DAOSPEC measurements and
``true" input EW on five artificial spectra with fixed R$=$35000 and S/N$=$100, and
different sampling ranging from 10 pixels (FWHM) to 1 pixel. As can be seen, all
cases with FWHM$\geq$2 pixels are comparable to each other, with the average
$\Delta$EW practically equal to zero and $\sigma\simeq$ 8--9~m\AA. In the case of
FWHM$=$1, the spread in $\Delta$EW is naturally much higher, while the average
$\Delta$EW remains reasonably close to zero. But a trend with EW seems present, in
the sense that smaller lines tend to be slightly overestimated in a regime of
under-sampling\footnote{This could be due to a larger uncertainty of the FWHM
estimate in a regime of undersampling.}. 

Since noise introduces variations at the pixel level, whenever the sampling is
close to the under-sampling limit (2 pixels), it is possible that DAOSPEC starts
confusing noise with features, thus increasing the variance of measurements as
in Figure~\ref{daopix} (bottom panel). Otherwise, the use of a correct first
guess of the FWHM improves the separation of noise
from real features.

To summarize, DAOSPEC produces reliable results for spectra that are adequately
sampled, but the possibility of using DAOSPEC to measure EW also in spectra that
are undersampled appears promising, although with a higher uncertainty
(especially for weak lines).  

\subsection{Performance considerations}
\label{dao-perf}

As an automatic program, DAOSPEC is fast if compared to interactive or
semi-automatic programs, especially when used in batch mode and without the
graphical interface. However, given the large computational power of modern
machines, and given that it can be run 24 hours a day, 7 days a week without
full-time supervision, DAOSPEC has been designed to give the most accurate results possible
rather than the fastest possible execution. In particular, the five big loops
described in Figure~\ref{flow} can take a long time in the interest of obtaining
more reliable continuum placement, FWHM estimate, and line parameters. 

A few tests on different machines show that one spectrum can be measured in a short
time, ranging from a few seconds to a few minutes, depending on the spectral quality
and other factors that are described in detailed in the following Sections. As an
example, DAOSPEC took a few minutes to measure 2429 line on the FEROS spectrum of
Section~\ref{sec-sou} (190\,000 pixels long), a Solar-type star with a crowded
spectrum and a complicated continuum shape. The above test was done on a Mac Pro
workstation, with a Dual-Core Intel Xeon Processor (2x2.66 GHz - 4 cores) and 2GB of
RAM memory, and the Mac OS X version 10.4.11 (Tiger). 

\subsubsection{Major factors}

The three major factors that have a large impact on the execution speed of DAOSPEC
are the following:

\begin{itemize}
\item{The MXSPEC parameter, within the main code file ({\tt daospec.f}), specifies the
maximum number of pixels in the spectra that will be analyzed. The higher this
value, the larger the amount of computer memory reserved. Setting MXSPEC to
unnecessarily high values can cause the code to go unbearably slow or even to
crash on some machines;}
\item{Setting the input FWHM to a severely wrong value can make the code go slower
by a factor of 10 or even 100, as many executions of the inner iteration loops attempt
to cope with the anomaly. We recommend deriving a robust estimate of FWHM
before starting the computation, as suggested in the DAOSPEC Cookbook
\citep{cookbook};}
\item{In general, the longer the spectrum (in pixels) and the larger the number of
lines in the spectrum (not in the laboratory line list), the longer it takes for
DAOSPEC to converge. In complicated spectra, setting a larger value for the smallest interesting EW
can reduce the number of loops and make the code go faster, although MXSPEC and
FWHM have a much stronger impact.}
\end{itemize}

\subsubsection{Secondary factors}

Other factors to take into consideration, although they have a smaller impact
\citep[see the DAO\-SPEC cookbook,][for more details]{cookbook}, are the following:

\begin{itemize}
\item{Adroit use of the Fortran compiler can optimize the code for faster execution;}
\item{Spectra with a simpler shapes, i.e., flatter continua, are faster to fit.
Similarly, using a low continuum order (if appropriate) makes the computation
faster, but it can also make execution significantly slower if used inappropriately (see
Section~\ref{sec-sou}). Also, setting the continuum ORDER=--1 (i.e., no continuum
fit) can also speed up the computation if the user is satisfied that the spectrum
has already been adequately rectified;}
\item{Using a FIXED FWHM makes the computation a bit faster, but please see
Figure~\ref{fwhm2} for the risks associated with a wrong FWHM estimate;}
\item{Some very small improvement can also be obtained if the fixed FWHM option is
used even for echelle spectra if they cover a very short wavelength interval,
but at the expense of the EW accuracy (Section~\ref{sec-fwhm});}
\item{Using a smaller interval for the radial velocity cross-correlation also makes
the computation faster. Leaving the search interval to its default value 
($\pm$500~km~s$^{-1}$) is only advisable if there is no prior constraint on the approximate
radial velocity of the objects;}
\end{itemize}

\section{Comparisons}
\label{sec-comp}

In this Section we compare results produced by DAOSPEC with data and programs from
the current literature. We summarize and discuss all the papers that, to our
knowledge, have made use of DAOSPEC in Section~\ref{sec-lett}, including our own
test with data from \citet{p02}. We also compare DAOSPEC with EWDET
(Section~\ref{sec-ewdet}) and with ARES (Section~\ref{sec-ares}) and we finally
perform an abundance analysis of the Sun (Section~\ref{sec-sun}).

\subsection{Literature tests on DAOSPEC}
\label{sec-lett}

DAOSPEC has been available to the astronomical community since 2002, when the
first test versions were circulated. Since then, it has evolved into the form
presented here and in the Cookbook \citep{cookbook}, and it has been used and
tested by some colleagues: a few authors used the code without mentioning any
specific tests \citep{mel03,p04,dall05a,dall05b,dall06,po05,zoc06,lec07}. 

Other papers compare DAOSPEC measurements to manual measurements with
IRAF, MIDAS or other methods and find a good agreement, but they do not
explicitly show the comparison
\citep{me04,das05,das06,bar06,bar07,let06,let07,ab06}. One paper used DAOSPEC to
measure radial velocities and mentions that a comparison with the results of
{\em fxcor} within IRAF gives agreement within the uncertainties
\citep{mon05}. Another couple of papers publish the comparison of EWs measured
by DAOSPEC with manual measurements \citep{ab05,vh05}; these are discussed
in Section~\ref{sec-tests}. 

To our knowledge, only two papers test DAO\-SPEC extensively: \citet{sousa} and
\citet{ares}, which are discussed in detail in the following sections.

\subsubsection{Basic Comparisons}
\label{sec-tests}

\citet{ab05} used DAOSPEC to measure EW of Fe~I and Fe~II lines in five red
giants in 47~Tuc. A comparison of IRAF interactive measurements with DAOSPEC
results on star 25 showed relatively good agreement, in the sense that the
average $\Delta$EW was smaller than the spread ($\sigma$$=$4.82~m\AA, see their
Figure~3). But a slight trend with EW appeared, in the sense that for larger EW
the $\Delta$EW became larger. According to the discussion in
Section~\ref{sec-fwhm} above, an effect like this could, for instance, be due
to a slightly ($\simeq$10\%) inappropriate input FWHM, but we do not know
whether something like that is in operation here.  In the end, the authors chose
to adopt the IRAF measurements to derive their atmospheric parameters and iron
abundances.  

Another, more favorable comparison was shown by \citet{vh05}, who plotted IRAF EW
measurements by \citet{shetrone} versus DAOSPEC, on GIRAFFE spectra
(R$\simeq$20000) of two stars in the Sculptor dwarf galaxy. They found good
agreement (within 10\%), with no sign of departures from the 1:1 relation for
strong lines up to 200~m\AA. This is expected if one considers the example of
Figure~\ref{fig-Gauss}, where we show that the Gaussian approximation is more and
more reliable, even for strong lines, as resolution goes down from
R$\simeq10^5$  to R$\simeq$20000.

\subsubsection{Detailed Comparisons}
\label{sec-sou}

Only two papers have performed detailed tests on DAOSPEC, namely \citet{sousa} and
\citet{ares}. 

\citet{sousa} used a synthetic (noiseless) model Solar spectrum of very high quality
(R$\simeq$120000 and S/N$\simeq$300) to compare DAOSPEC and IRAF EW measurements. 
They found essentially perfect agreement in a red window (6000--6300\AA) with
$\Delta$EW$=$0.8$\pm$1.1 m\AA, based on 34 lines, and fair agreement in a blue
window (4400--4650\AA) with $\Delta$EW$=$4.0$\pm$4.9 m\AA, based on 25 lines (see
also Figure~\ref{fig-ares}). This must of course be due to the higher crowding level
and lower S/N ratio of the blue part of the spectrum.  

The synthetic spectrum was then degraded both in S/N ratio and resolution, and
the DAOSPEC measurements were compared with each other. DAOSPEC appears to give
very different average EWs and variance, by as much as
$\Delta$EW=15$\pm$20~m\AA, for the lowest resolution case (S/N$\simeq$10 and
R$\simeq$12000). While some increase in the variance can be easily understood
when varying S/N ratio or resolution, as can be seen in our own tests
(Section~\ref{dao-qual}), EW discrepancies and variance as large as those
reported by \citet{sousa} are difficult to understand, and indeed we do not find
such behavior in our own tests (Sections~\ref{sec-sn} and \ref{sec-res}).  

\begin{figure}[t]
\includegraphics[width=5.5cm, angle=270]{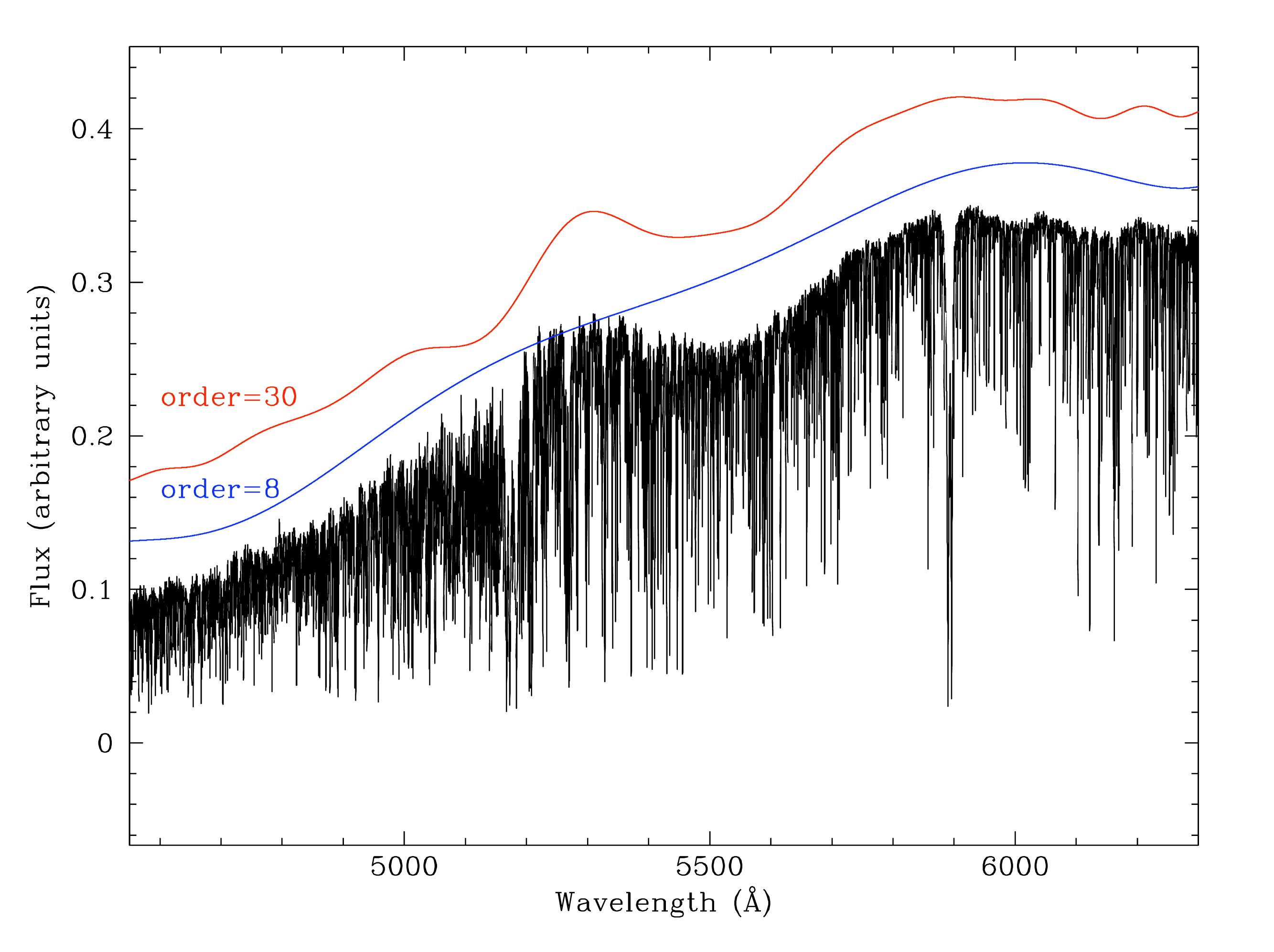}
\caption{A typical FEROS spectrum of a Solar-type star (kindly provided by S.
Sousa). Two Legendre DAOSPEC polynomials are overplotted, with an arbitrary
vertical offset for clarity. As can be seen, the Legendre polynomial of 8$^{th}$
order (lower continuous line) does not adequately represent the spectral shape,
while the 30$^{th}$ order polynomial (upper continuous line) fits the spectrum 
better. \label{fig-feros}}
\end{figure}

When measuring their FEROS spectra, \citet{sousa} encountered some problems. In
particular, they found an enormous, unacceptable spread in the resulting EWs, and
they managed to obtain reasonable EWs only by cutting the spectrum into 100~\AA\ 
segments and running DAOSPEC manually on each small piece. The $\Delta$EW
from IRAF measurements then went down from 12.1$\pm$17.1~m\AA\  to 3.0$\pm$ 4.7~m\AA\ 
(S. Sousa, private communication) but, of course, at the expense of execution time
and humanpower (several hours). They kindly provided us with some of their FEROS
spectra (Figure~\ref{fig-feros}) and we have repeated their measurements. We found that
using a different order for the continuum fit (30 instead of 8,
Figure~\ref{fig-feros}) and a different FWHM (14 instead of 5,
Section~\ref{sec-fwhm}, Figures~\ref{fwhm1} and \ref{fwhm2}) gave much better
EWs and decreased the execution time by a factor of 50, roughly. We also tried
cutting the spectrum into short pieces, both as a consistency check and to test the
execution times, but we used shell scripts \citep[Cookbook,][]{cookbook} to run
DAOSPEC automatically---in 10 minutes, total time (see Section~\ref{dao-perf})---on the various
pieces: we obtained $\Delta$EW$=$--4.1$\pm$4.3~m\AA\ when using the full spectrum,
and $\Delta$EW$=$--6.5$\pm$4.4~m\AA\ when the spectrum was cut into 100~\AA\ 
pieces.

The paper that introduced ARES \citep{ares} was the second to perform a detailed
check on DAOSPEC, using the same datasets as \citet{sousa} and the same DAOSPEC
configuration parameters. ARES is based on the IRAF task {\em splot}, and therefore
the first comparisons made were between ARES and IRAF, and between DAOSPEC and IRAF. The
results obtained with ARES were more similar to IRAF than the ones with DAOSPEC,
supporting the conclusion that ARES is a very well designed extension of {\em
splot}. We have seen, however, that the most important factor in these
comparisons can be the way the continuum is chosen. In case of crowded spectra,
we have claimed that the algorithm employed by DAOSPEC can give better results
(Section~\ref{sec-arti}), but since both IRAF and ARES are highly customizable in
terms of continuum placement, we do not doubt experienced and careful users
can obtain good results with those algorithms.  

To summarize, an appropriate choice of the configuration parameters is crucial to
obtain good results with DAOSPEC. The Cookbook \citep{cookbook} provides practical
and objective methods for finding the best values for these parameters, as well as the
discussions and tests presented in Section~\ref{dao-pars} here.   

\subsubsection{Red Giants in $\omega$~Centauri}
\label{sec-p02}

The dataset of EW measurements that \cite{p02} obtained with the IRAF task {\em splot} to derive
abundances for six red giant stars in $\omega$~Cen constitutes a good testbed for DAOSPEC. The full data description
can be found in the original paper; in short, the six spectra were taken
with UVES at the Very Large Telescope in Paranal, Chile, with $R\simeq$45000 and
S/N$\simeq$100--150 per resolution element, covering the range 5250--6920\AA.
Stellar metallicities range from \hbox{[Fe/H]=--0.49} to \hbox{--1.20}, with temperatures
around 4000~K and gravities of about 1~dex. The input line list contains 230
features of various elements, although only [Fe/H], [Ca/Fe], [Si/Fe] and [Cu/Fe]
were published by \citet{p02}. 

\begin{figure}[t]
\includegraphics[angle=270,scale=0.30]{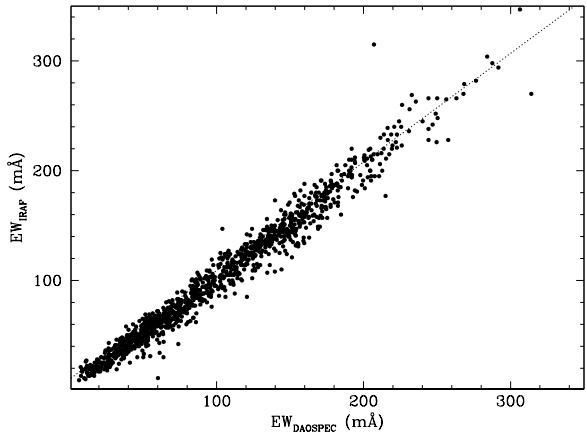}
\caption{Comparison of the original measurements from \citet{p02},
obtained with IRAF (Y axis), and the measurements obtained here with
DAOSPEC with the same line list and on the same spectra (X axis). Perfect
agreement is marked with a dotted line. \label{p02fig}}
\end{figure}

We remeasured these spectra with DAOSPEC and compared the results
(Figure~\ref{p02fig}). A total of 1150 lines were used in the comparison. We
found a very good average agreement, with DAOSPEC measurements marginally smaller, by
$\Delta$EW$=$--1.3$\pm$10.3~m\AA. When considering the six stars separately, we
found differences ranging from $\Delta$EW$=$--3.7$\pm$10.7~m\AA, for star
WFI~222068, which is the most metal rich of the sample, to
$\Delta$EW$=$1.1$\pm$7.1~m\AA, for star WFI~618854, which is the most metal-poor of
the sample. No trend with EW is apparent. 

The agreement appears satisfactory within the uncertainties, expecially in the
light of the tests performed in Section~\ref{sec-arti}, where we show again that
an agreement between DAOSPEC and IRAF measurements gets naturally worse as
metallicity (and line crowding) increases.

\subsection{DAOSPEC vs. EWDET}
\label{sec-ewdet}

\begin{figure}[t]
\includegraphics[angle=270,scale=0.30]{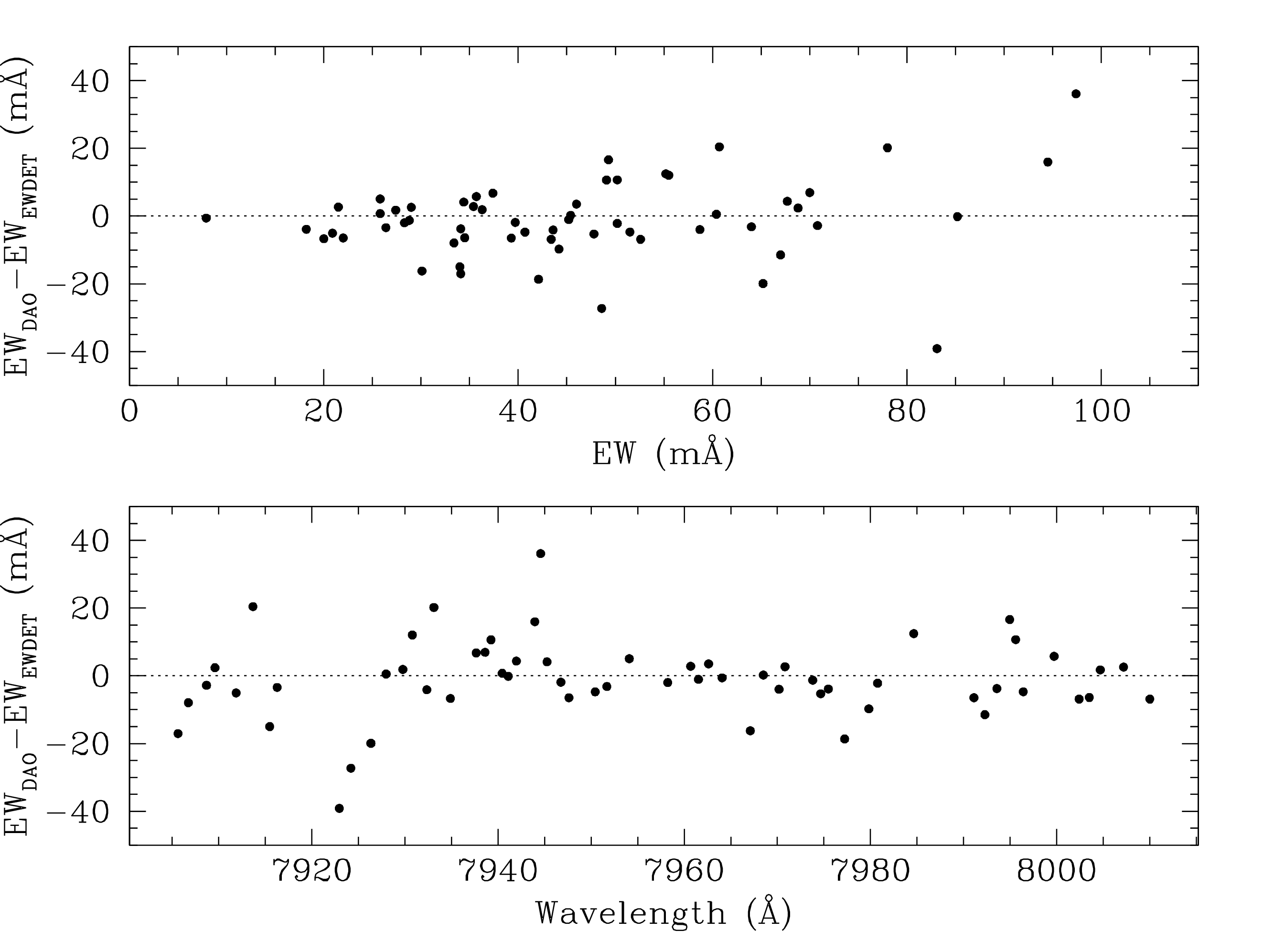}
\caption{Difference between DAOSPEC and EWDET EWs measured on a test spectrum of
a giant in M~71 provided with EWDET. $\Delta$EW is plotted versus EW (top panel)
and versus wavelength (bottom panel). The average difference is
$\Delta$EW$=$--1.2$\pm$11.7~m\AA. \label{fig-ewdet}}
\end{figure}

EWDET \citep[][see also Section~\ref{sec-ew}]{ewdet} was obtained by courtesy of
S. Ram\i\'rez.  It came with a test spectrum of a moderately crowded red giant
in M~71, covering the range 7900--8000~\AA, with R$\simeq$30000 and a S/N ratio of at least
100 everywhere. We used this spectrum with the default configuration file ewdet.inp
provided with the code, to measure the EWs of 70 lines. Ten additional lines
were found, but EWDET did not report an EW for any of them because the Gaussian
fit did not converge. All the lines found by EWDET were used as the input
``laboratory'' line list for DAOSPEC, which we then used to obtain EWs from the
same spectrum.  It is perhaps worth stressing here that the input line list
plays no part in the {\it finding\/} of candidate spectral lines by DAOSPEC; it
is only after features have been detected that tentative {\it identifications\/}
with features in the input list are sought.  There is no attempt to ``force'' the
detection of features in the spectrum at wavelengths specified by the input
laboratory list.  In the present case, DAOSPEC was able to (independently) find
and measure all the lines that EWDET had found, including the 10 that EWDET
had subsequently discarded. No apparent defect was found on those 10 lines in a
visual inspection of the spectrum.  

Figure~\ref{fig-ewdet} plots the difference between the DAOSPEC and EWDET EWs
versus EW (top panel) and versus wavelength (bottom panel). The average
difference is $\Delta$EW$=$--1.2\AA\ with a variance of 11.7~m\AA. While the
two sets of measurements appear in good agreement, the spread is slightly higher
than expected, i.e., higher than that found in the comparison of
Section~\ref{sec-p02} between DAOSPEC and hand-made IRAF measurements. A trend
of increasing spread with increasing EW might be present, while no obvious trend
with wavelength is seen.

A comparison of the model continua adopted by the two programs shows a an overall
systematic difference of 1.3\% (DAOSPEC continuum lower), with a variance of 0.6\%
around this mean offset.  For comparison, the residual spectrum produced by
DAOSPEC has a pixel-to-pixel flux variance of 2\%.  On the surface, this case
appears to be similar to that discussed in Section~\ref{sec-con} above.  Such a
discrepancy in the continuum levels could be the cause of the small $\Delta$EW
offset found between the two codes (see Section~\ref{sec-ew},
Figure~\ref{fig-levs}).  

On average, the standard errors estimated by DAOSPEC are larger by
0.7$\pm$1.5~m\AA\  than those reported by EWDET, even though the latter also
includes the uncertainty due to the continuum placement and the former does not.
In any case, given the large spread in $\Delta$EW seen in
Figure~\ref{fig-ewdet}, both error estimates appear a bit small, indicating that
some other unidentified source of uncertainty might be present. If we estimate
an error budget, including the average errors by EWDET ($\sim$3~m\AA) and
DAOSPEC ($\sim$4~m\AA) and an error due to the continuum placement as estimated
roughly from Figure~\ref{fig-levs} ($\sim$7~m\AA), we account for a spread of
$\sim$9~m\AA, i.e., the missing source of uncertainty must be of about
$\sim$8~m\AA\footnote{One thing to note here is that the spectral range chosen
includes a relatively strong H$_2$O telluric absorption band, which could cause
some of the observed scatter.}. This might suggest that the EWDET continuum
placement uncertainty \citep[][and Section~\ref{placement} here]{ewdet} might be
underestimated. 

Finally the average difference between the FWHM found by EWDET for each line,
and the FWHM found by DAOSPEC (scaled with wavelength) is
$\Delta$FWHM$=$0.001$\pm$0.076~\AA, and the average radial velocity difference
between the two sets, in the sense EWDET minus DAOSPEC, of measurements is very
small, $\Delta v_r$$=$0.1$\pm$0.6~km~s$^{-1}$. 

Summarizing, the comparison can be considered satisfactory once all the sources of
uncertainty are properly taken into account. The only minor disadvantage of EWDET is
the fact that it has been written for personal use and it requires knowledge of
Fortran to manually adapt some routines to meet the needs of each set of spectra,
including naming conventions and so on.

\subsection{DAOSPEC vs. ARES}
\label{sec-ares}

\begin{figure}[t]
\includegraphics[angle=270,scale=0.30]{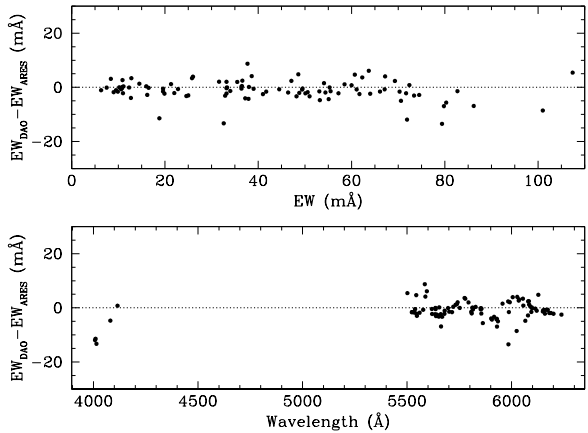}
\caption{Difference between DAOSPEC and ARES EWs measured on a test spectrum of
the Sun provided with ARES. $\Delta$EW is plotted versus EW (top panel) and
versus wavelength (bottom panel). The average difference is
$\Delta$EW$=$--1.1$\pm$3.7~m\AA. \label{fig-ares}}
\end{figure}

ARES \citep{ares} has been obtained from the ARES
webpage\footnote{http://www.astro.up.pt/$\sim$sousasag/ares/} with, {\it inter
alia\/}, a test spectrum of the Sun obtained with HARPS from observations of
Ganymede\footnote{http://www.ls.eso.org/lasilla/sciops/3p6/harps/monitorin
g/sun.html}. Similarly to what we have done with EWDET, we ran ARES on the test
spectrum and used its output as an input line list for DAOSPEC. 

\begin{deluxetable}{lcccccc}
\tablecaption{Abundance Analysis of the Sun \label{tab-sun}}
\tablewidth{0pt}
\tablehead{\colhead{EW} & \colhead{T$_{\rm{eff}}$} & \colhead{$\log g$} 
& \colhead{$v_t$} & \colhead{[FeI/H]} & \colhead{FeII/H]} & $n_{\rm{lines}}$ \\ 
& (K) & (dex) & (km/s) & (dex) & (dex) &}
\startdata
\citet{moore}  & 5780 & 4.46 & 0.7 & -0.012$\pm$0.005 & +0.095$\pm$0.012 & 160 \\  
\citet{rutten} & 5780 & 4.43 & 0.8 & -0.019$\pm$0.005 & +0.010$\pm$0.013 & 197 \\
DAOSPEC        & 5780 & 4.44 & 0.6 & +0.005$\pm$0.008 & -0.023$\pm$0.018 & 195 \\
\enddata
\end{deluxetable}

The average differences of key parameters, in the sense of DAOSPEC minus ARES,
can be summarized as follows: $\Delta$EW$=$--1.1$\pm$3.7~m\AA,
$\Delta$FWHM$=$0.01$\pm$0.05~\AA\footnote{This difference is 100 times larger
than in the case of the EWDET comparison. However, it corresponds to a
$\simeq$8\% relative difference, which, according to our Figure~\ref{fwhm2},
implies a $\Delta$EW of a few m\AA\  at most, compatible with the quoted
$\Delta$EW$=$--1.1$\pm$3.7~m\AA.}, $\Delta v_r$$=$--0.002$\pm$0.126~km~s$^{-1}$,
based on 98 lines in common. No error estimate is provided by ARES. At first
glance, all these values appear in very good agreement within the uncertainties,
even better than the comparison made with EWDET in Section~\ref{sec-ewdet}. This
is especially true when considering the spread in $\Delta$EW, which is
11.7~m\AA\  in the comparison with EWDET and only  3.7~m\AA\ in the comparison
with ARES. The very good agreement must of course be largely due to the fine
quality of the test spectrum, which has R$\simeq$45000 and S/N$\simeq$350.
Figure~\ref{fig-ares} confirms good agreement with no trends with wavelength or
EW in the differences, except for a possible problem in the bluest and most
crowded part of the spectrum. A last comparison was made on the number of lines
found. The authors do not mention how many lines were found and/or identified by
each code, but state that ARES finds more lines than DAOSPEC. If we compare the
Solar spectra taken from Ganymede, we find that ARES identifies 101 lines, and
DAOSPEC identifies 100. 

ARES and DAOSPEC represent two very different ways of approaching the problem of
measuring EWs. ARES closely follows IRAF, including a major IRAF feature,
namely the possibility to customize the continuum fitting procedure. Because of the
way the continuum is fit, ARES is faster than DAOSPEC, although maybe a bit longer
to configure. ARES takes of the order of seconds for each spectrum, while DAOSPEC
may take from a few seconds to a few minutes, depending on the spectrum
characteristics. Finally ARES gives no error estimate or radial velocity; indeed, 
the radial velocity is one of the necessary inputs, not outputs of the code. 

Nevertheless, in spite of the different continuum placement philosophies, ARES and
DAOSPEC give entirely comparable measurements, within the uncertainties.

\subsection{Abundance analysis of the Sun}
\label{sec-sun}

As a final test, DAOSPEC was used on the Solar spectrum obtained with HARPS
(Section~\ref{sec-ares}) to derive iron abundances for the Sun. The results have
been compared, using the same models and abundance calculation code, to the
abundances obtained with the EWs measured by \citet{moore} and \citet{rutten}.

To measure EWs with DAOSPEC, we created a line list containing all lines in common
between \citet{moore} and \citet{rutten}. This line list was fed to DAOSPEC and,
for homogeneity, it was also used to derive the Solar abundance with the
original \citet{moore} and \citet{rutten} measurements. 

We used the atmospheric models by \citet{edv} and the latest version of the
abundance calculation code originally published by \citet{spite}. For sake of
homogeneity, the atomic parameters (including log~$gf$) were taken from the
line list of \citet{rutten}. In this way, the only difference among the three
analyses comes from the EWs. The Solar temperature was kept fixed at 5780~K;
gravity was allowed to vary between log~$g$=4.4 and 4.5, to allow for
micro-adjustments of the Fe~I and Fe~II ionization equilibrium; the
micro-turbulent velocity was kept as a free parameter and the best value was
chosen as the one that minimized the slope of the EW vs. [Fe/H] relation. 

The results of our analysis are shown in Table~\ref{tab-sun}, where it can
clearly be seen that only negligible variations in $log$ g and very small
variations of the microturbulence ($v_t$) were necessary to obtain abundances
that are quite compatible with each other, and with the Solar values.

\section{Conclusions}
\label{sec-dis}

While we have argued that one of the advantages of DAOSPEC is that it is
minimally interactive, and hence fast and reproducible, we must confess
that there still remain some user-specified reduction options whose best values
are not entirely obvious.  Examples of these are the polynomial order to be used
in the continuum fits, the initial estimate of the FWHM value appropriate to a
given spectrum, the range of plausible radial-velocity values to consider, and
the minimum EW that is considered to be of interest.  In some sense, however, we
feel that this evident weakness is also a not-so-evident strength: the program
runs sufficiently fast that the user is easily able to alter each of the various
input parameters over some reasonable range of values and see directly how the
astrophysical results are changed when all other inputs are held strictly fixed.
We have also explored in detail the behavior of DAOSPEC in several cases of
interest, performing a variety of tests that show the range of applicability of
DAOSPEC in terms of data quality, and should also help in the search for the best
set of configuration parameters for each case. Novice (and experienced!) users
are positively encouraged to carry out such experiments. Similarly, feedback from
the users is of immense help, allowing us to refine the code in the directions of
major interest for the scientific community.

\acknowledgments

We would like to thank the early testers and users and a few people that gave us
useful suggestions and stimulated interesting discussions: Stefano Bagnulo, Katia
Biazzo, Giulia Caira, Eugenio Carretta, \'Eric Depagne, Patrick Fran\c cois,
Raffaele Gratton, Vanessa Hill, Segei A. Levshakov, Livia Origlia, Luca Pasquini,
Ignazio Pillitteri, Francesca Primas, Marina Rejkuba, Emanuel Rossetti, Luca
Sbordone, Antonio Sollima, Manuela Zoccali. We thank Elizabeth Griffin for a
critical reading of the manuscript. We would also like to thank Chris Sneden,
Solange Ram\'{\i}rez and Sergio Sousa for providing their codes and their help. We
ackowledge the European Southern Observatory in Garching, Germany, where part of
this work has been carried out, through the ESO {\em Studentship} and {\em Science
Visitorship} programmes. E.~P. would like to thank the Dominion Astronomical
Observatory in Victoria, Canada, for the warm hospitality, and to acknowledge the
financial support of the INAF, Osservatorio Astronomico di Bologna, throughout the
project.

\end{document}